\documentclass[12pt]{article}

\usepackage{amsmath, amsthm, mathrsfs, amssymb, amsfonts}
\usepackage{soul}
\usepackage{dsfont}
\usepackage[authoryear]{natbib}
\usepackage{times, latexsym}
\usepackage{xcolor}
\usepackage[table]{xcolor} 
\definecolor{shade}{gray}{0.90} 
\usepackage{float}
\usepackage{setspace}
\usepackage{graphicx, epsfig}
\usepackage{subfigure}
\DeclareGraphicsExtensions{.eps}
\usepackage{bbm}
\usepackage{mathtools}
\usepackage{framed}
\usepackage{fancybox}
\usepackage{rotating, threeparttable, booktabs, caption, dcolumn, pdflscape, enumitem}
\usepackage{arydshln}
\usepackage{multirow} 
\usepackage[titletoc, title]{appendix}
\newcolumntype{d}[1]{D..{#1}}

\usepackage[
					colorlinks=true,                     
            		 	citecolor=blue,                       
           			linkcolor=blue,                      
            			urlcolor=blue,                        
            		filecolor=blue,                       
           			bookmarksnumbered=true,    
           			bookmarksopen=true,            
           			breaklinks=true,                     
            		 	pdfstartview=FitH,                 
]{hyperref}

\usepackage{algorithm}
\usepackage{algpseudocode}

\algnewcommand\algorithmicassume{\textbf{Assume:}}
\algnewcommand\Assume{\item[\algorithmicassume]}
\algnewcommand{\LineComment}[1]{\State \(\triangleright\) #1}

			\def\calD{\mathcal{D}}

\renewcommand{\tilde}{\widetilde}

\def\prob {{\rm Pr}}

\def\argmin{\mbox{\rm argmin}}
\def\argmax{\mbox{\rm argmax}}

\allowdisplaybreaks[4]
\usepackage{listings}

\title{Shared Keyboard: An improved Bayesian design for phase I clinical trials via Beta kernel process}

\author{Jiangyan Zhao$^{1}$,  Xian Shi$^{1}$,  and Jin Xu$^{1,2,}$\footnote{Corresponding author: School of Statistics, East China Normal University, 3663 North Zhongshan Road, Shanghai 200062, China, email: {{jxu@stat.ecnu.edu.cn}}}}

\date{}

\begin{document}

\maketitle

\vskip -1cm

\noindent $^1$School of Statistics, East China Normal University, Shanghai, China\\
$^2$Key Laboratory of Advanced Theory and Application in Statistics and Data Science -- MOE, East China Normal University, Shanghai, China




\begin{abstract}
Model-assisted interval designs such as the Keyboard design are transparent and easy to implement in phase I oncology trials. However, interim decisions based solely on data from the current dose may overlook informative signals from neighbouring doses, leading to unnecessary escalation or de-escalation. We propose the shared Keyboard design, a Bayesian model-assisted design that replaces the independent beta--binomial updating scheme at each dose with a posterior induced by a Beta kernel process using kernel-weighted pseudo-counts. The design preserves the decision structure of the Keyboard design while enabling controlled borrowing across nearby doses. To prioritise overdose control, we propose an asymmetric kernel that assigns greater weight to toxicities observed at higher doses during escalation. We further extend the proposed design to accommodate adaptive dose insertion when the initial dose grid is inadequate and time-to-event outcomes when late-onset toxicities are present. Extensive simulation studies demonstrate substantial improvements in both accuracy and safety for identifying the maximum tolerated dose. In settings involving dose insertion, the proposed design identifies inserted target doses more effectively than adaptive dose modification while maintaining a comparable modification rate. 
\end{abstract}

\noindent \textbf{Keywords:} Beta kernel process; dose finding; dose insertion; Keyboard design; model-assisted design; phase I clinical trials

\newpage

\section{Introduction}

Phase I oncology trials aim to identify a dose with an acceptable toxicity profile for late clinical development. In the conventional cytotoxic setting, this objective is to identify the maximum tolerated dose (MTD) whose dose-limiting toxicity (DLT) probability is closest to a prespecified target rate  \citep{LeTourneau2009, Yuan2023BOIN}. 

The traditional 3+3 design \citep{Storer1989} is widely used for its simplicity and ease of implementation. However, it suffers from low accuracy and inefficiency \citep{Reiner1999StandardUpDown, LeTourneau2009}, particularly as modern oncology trials demand higher success rates and greater statistical rigour.
Bayesian methods have become a natural alternative and now play a central role in modern phase I trial design \citep{Ji2026Bayesian, FDA2026Bayesian}.
Broadly speaking, Bayesian phase I designs may be divided into model-based and model-assisted approaches. 
Model-based designs rely on an explicit parametric model for the dose--toxicity relationship, whereas model-assisted designs use Bayesian decision rules that retain transparency and ease of implementation.
The continual reassessment method (CRM) \citep{OQuigley1990CRM} is a well known model-based design that uses all accumulated data to estimate the dose--toxicity relationship and has been shown to outperform rule-based methods in many settings \citep{Iasonos2008CRMvs3plus3}. 
However, CRM requires subjective specification of a parametric model and a skeleton of toxicity rates. It suffers from model misspecification and overdose risk \citep{Yin2009BMACRM, Zhou2018CompMTD}.

To combine the simplicity and transparency of rule-based designs with the statistical efficiency of model-based designs, model-assisted interval designs have become increasingly popular. 
The Bayesian optimal interval (BOIN) design \citep{Liu2015BOIN} uses calibrated toxicity intervals to guide escalation and de-escalation and is straightforward to implement. 
The Keyboard design \citep{Yan2017Keyboard} is another model-assisted design that partitions the toxicity scale into a set of equal-width intervals (keys), identifies the strongest key under a beta--binomial posterior, and then determines whether to escalate, de-escalate, or stay. 
Compared with BOIN, the Keyboard design retains the operational simplicity of interval-based rules while incorporating posterior uncertainty more explicitly. 
As a result, BOIN and Keyboard designs are now among the most widely used model-assisted methods in early-phase oncology trials \citep{Yuan2023BOIN, Ananthakrishnan2022OverviewBOIN}.

Despite their practical success, interval-based designs share an important limitation: interim decisions are driven primarily by data observed at the current dose level. 
This current-dose-alone strategy ignores potentially relevant information from nearby doses that have already been explored. 
In small-sample settings, it can lead to unstable posterior learning, delayed reaction to emerging toxicity signals at adjacent higher doses, and non-monotone dose-wise toxicity estimates \citep{Lin2019CION, Ananthakrishnan2022OverviewBOIN}. 
Although isotonic regression is commonly used at the end of the trial to enforce monotonicity, it is only a post hoc adjustment and does not address the information inefficiency of local-only interim updating. 
This limitation is especially relevant, where neighbouring doses are often clinically close and cautious local borrowing may improve both safety and MTD selection.

A second practical limitation is that the prespecified dose grid may be inadequate \citep{Hu2013ADI, Chu2016ADM}. 
In many trials, the available dose levels are sparse, so that the target toxicity region may lie between two adjacent prespecified doses or even outside the explored range. 
Adaptive dose-insertion methods have therefore been proposed to refine the grid during the trial.  
\citet{Hu2013ADI} proposed a relatively unified model-based insertion framework, where dose insertion is triggered by an activation rule and the new dose is determined through a parametric dose--toxicity model.  
\citet{Chu2016ADM} proposed a more robust and decoupled strategy, in which the triggering criterion is evaluated using beta--binomial posterior probabilities whereas the new dose is estimated separately through a nonparametric local polynomial model \citep{Fan1996LocalModel}. 
Both methods are subject to model misspecification.

In this article, we propose the \emph{shared Keyboard design} (SKBD), an extension of the Keyboard design that preserves its key-based decision structure while addressing both limitations described above. 
The central idea is to replace the independent beta--binomial posterior used at each dose by a Beta kernel process (BKP)-induced posterior \citep{Goetschalckx2011CCBP, Zhao2025BKP}, so that inference at the current dose can borrow information from neighbouring doses through kernel-weighted pseudo-counts. In particular, the proposed design adopts an asymmetric borrowing scheme to give priority to overdose control. 
Second, the same BKP-based framework also supports adaptive dose insertion, leading to the \emph{insertion shared Keyboard design} (Ins-SKBD), when the prespecified dose grid is inadequate. 
Unlike parametric inverse-model approach or decoupled modification approach, candidate refinement locations are evaluated through the same kernel-weighted posterior updating principle, so that dose transition and dose-grid refinement are handled within a coherent Bayesian model-assisted framework. 
Third, we further extend the same framework to handle the case with late-onset toxicity \citep{Zhou2024TITE} through a \emph{time-to-event shared Keyboard design} (TITE-SKBD). 

Simulation studies show that SKBD improves MTD selection and overdose control over Keyboard design. 
For scenarios in which grid refinement is needed, Ins-SKBD more effectively identifies target inserted doses than a competing adaptive dose-modification approach while maintaining a comparable modification rate. 
An R package, \texttt{SKBD}, is publicly available on GitHub at \url{https://github.com/Jiangyan-Zhao/SKBD}, together with a Shiny-based graphical interface to facilitate implementation and trial planning. 

The remainder of this article is organised as follows. 
Section~\ref{sec:method} reviews the Keyboard design and presents SKBD, Ins-SKBD, and TITE-SKBD. 
Section~\ref{sec:simulation} evaluates the operating characteristics of SKBD and its extensions through simulation studies. 
Section~\ref{sec:discussion} concludes with a discussion of practical implications, limitations, and possible future extensions.

\section{Methods}\label{sec:method}

\subsection{Notation}\label{sec:notation}

Consider $J$ prespecified dose levels for an investigational agent, sorted as $\mathcal{J}=\{d_1<\cdots<d_J\}$.
Let $\pi(d)=\Pr(\text{DLT}|d)$ denote the dose-limiting toxicity (DLT) probability at dose $d$,
and assume that $\pi(d)$ is non-decreasing in $d$.
Write $\pi_j=\pi(d_j)$ for $j=1,\ldots,J$.
Let $\phi\in(0,1)$ denote the target DLT rate, which is usually in 20--33\% \citep{Wages2018CRMweb}. The maximum tolerated dose (MTD) is defined as the dose in $\mathcal{J}$ whose toxicity probability is closest to $\phi$,
i.e., $d_{j^*}$ with $j^*=\arg\min_{1\le j\le J}|\pi_j-\phi|$. 
(If ties occur, select the lowest level as a conservative convention.)

\subsection{Keyboard Design}\label{sec:KBD}

We briefly review the keyboard design \citep{Yan2017Keyboard}, which serves as the decision framework for our improved extension in Section~\ref{sec:SKBD}.

{\it (i) Definition of keys} 
In the range of toxicity probability, define the \emph{target key} (target toxicity interval) as
$\mathcal{I}_{\text{target}} = (\phi - \varepsilon_1,\ \phi + \varepsilon_2)$,
where $\varepsilon_1>0$ and $\varepsilon_2>0$ are clinically tolerable deviations from $\phi$.
A common choice is $\varepsilon_1=\varepsilon_2=0.05$, yielding a target key of width $0.1$.
Any dose whose toxicity probability lies in $\mathcal{I}_{\text{target}}$ is regarded as a plausible MTD.

The keyboard design partitions the unit interval $(0,1)$ into a collection of adjacent toxicity intervals (``keys'') that include the target key and cover the full range of plausible DLT rates.
Denote these keys by $\mathcal{I}_1,\mathcal{I}_2,\ldots,\mathcal{I}_L$, with $\mathcal{I}_{\ell_t}=\mathcal{I}_{\text{target}}$.
For example, if $\phi = 0.3$ and $\mathcal{I}_{\text{target}} = (0.25, 0.35)$, then a typical set of keys is
$(0,0.05)$, $(0.05, 0.15)$, $(0.15, 0.25)$, \ldots, $(0.85, 0.95)$, $(0.95, 1)$, where the boundary keys are truncated by 0 and 1.
Clearly, the number of keys $L$ depends on the key width and the value of $\phi$.

{\it (ii) Identify the strongest key to guide dose escalation} 
Suppose that at the current dose level $j$, a total of $n_j$ patients have been treated and $y_j$ of them experienced DLTs.
Let $\mathcal{D}_j=\{(n_j,y_j)\}$ denote the data accumulated at the \emph{current dose level}.
The keyboard design posits a beta--binomial model $\pi_j \sim \textrm{Beta}(a_j, b_j)$ and $y_j\sim \textrm{Bin}(n_j, \pi_j)$,
which yields the conjugate posterior
\begin{equation}\label{eq:BetaPost}
	\pi_j\mid\calD_j\sim\textrm{Beta}(a_j+y_j, b_j+n_j-y_j).
\end{equation}
For each key $\mathcal{I}_{\ell}$ ($\ell=1,\ldots,L$), compute the posterior key probability $\prob(\pi_j \in \mathcal{I}_{\ell} \mid \mathcal{D}_j)$.
The \emph{strongest key} is defined as the key with the largest posterior probability, with index $\ell_{\rm s} = \argmax_{1\leq\ell\leq L} \{\prob(\pi_j \in \mathcal{I}_{\ell} \mid \mathcal{D}_j)\}$. 
If ties occur, a practical conservative tie-breaking rule is to choose the \emph{higher} key (i.e., the more toxic interval),
which avoids overly optimistic escalation driven by ties.

Dose transition is determined by comparing the strongest key with the target key:
\[\begin{dcases}
	\textrm{escalate}, & \textrm{if}\quad \ell_{\rm s} < \ell_{\rm t},  \\
	\textrm{stay}, & \textrm{if}\quad \ell_{\rm s} = \ell_{\rm t},  \\
	\textrm{de-escalate}, & \textrm{if}\quad \ell_{\rm s} > \ell_{\rm t}.
\end{dcases}\]
The rationale is clear. The strongest key reflects the toxicity range most supported by the interim posterior, guiding escalation if below the target, de-escalation if above it, and continuation if aligned with it.

The above procedure is repeated until the maximum sample size is exhausted or an early-stopping rule is triggered.
A standard rule for overdose-control in the keyboard design is to eliminate dose $d_j$ and the other higher doses if $\prob(\pi_j > \phi \mid \mathcal{D}_j) > 0.95$ and $n_j\geq 3$ \citep{Yan2017Keyboard, Yuan2023BOIN}. If the lowest dose is eliminated, the trial is terminated for safety.

{\it (iii) Select the MTD after isotonic regression adjustment} 
At the end of the trial, isotonic regression is applied to the observed DLT rates
$\widehat{\pi}_j=y_j/n_j$ to enforce the monotonic dose-toxicity relationship.
In particular, the pooled adjacent violators algorithm (PAVA) \citep{Barlow1972PAVA} is implemented to yield a non-decreasing sequence
$\widetilde{\pi}_1\le \cdots \le \widetilde{\pi}_J$. Finally, the estimated MTD is selected at dose level $\widehat{j}^*=\argmin_{1\le j\le J}\ |\widetilde{\pi}_j-\phi|$.

This post hoc adjustment is routinely used because the keyboard design makes local, rule-based decisions and the per-dose sample sizes are often limited, so the observed rates $\widehat{\pi}_j$ may not hold monotonicity automatically. 

\subsection{Shared Keyboard Design}\label{sec:SKBD}

As pointed before, the keyboard design makes interim decisions on \emph{local} information alone through \eqref{eq:BetaPost} and may overlook the emerging toxicity signals at neighbouring doses. We propose a shared keyboard design (SKBD) to overcome this deficiency. The idea is to replace the independent beta--binomial posterior with an information-sharing composition based on the beta kernel process (BKP) \citep{Goetschalckx2011CCBP, Zhao2025BKP}.

Let $\mathcal{D}=\{(y_j, n_j): j=1,\ldots,J\}$ denote the cumulative trial data, and let $\mathcal S=\{s: n_s>0\}$ denote the set of dose levels with observed outcomes.
Define the kernel-weighted pseudo-counts and sample size by 
\begin{equation}\label{eq:pseudo-counts}
y'_j=\sum_{s\in\mathcal S} w(d_j;d_s)\,y_s,\quad 
n'_j=\sum_{s\in\mathcal S} w(d_j;d_s)\,n_s,
\end{equation}
where the normalised weights are given by
\[
	w(d; d_s)=\frac{k(d, d_s)}{\sum_{u\in\mathcal S} k(d, d_u)}, \qquad s\in\mathcal S.
\]
and $k(d,d')\ge 0$ is a kernel function that quantifies the similarity between two dose levels and thereby determines how information is borrowed across doses. The specific kernel form and the choice of kernel scale parameters are described in Section~\ref{sec:Kernel}. (Note that weighted counts $y'_j$ and sample size $n'_j$ need not be integers.)
Because the weights are normalised over $\mathcal S$,  $n'_j$ is a kernel-weighted average of the accrued dose-level sample sizes. This normalisation prevents the posterior precision from increasing with the number of observed dose levels and gives $w(d_j;d_s)$ a direct interpretation as the relative contribution of dose $d_s$ to inference at $d_j$. 

Then, SKBD defines a BKP-induced posterior distribution of $\pi_j$ as
\begin{equation}\label{eq:BKPmodel}
	\pi_j \mid \mathcal{D} \sim \mathrm{Beta}\!
\left(a_j+y'_j,\; b_j+n'_j-y'_j\right).
\end{equation}
This representation makes the role of borrowing explicit. In particular, the posterior mean of $\pi_j$ is driven by the kernel-weighted DLT rate $y'_j/n'_j$ together with the prior specification, whereas the posterior precision is determined by the total amount of borrowed information, namely $a_j+b_j+n'_j$. 
We set $a_j=b_j\equiv 1$ for all $j$ by default, which corresponds to a non-informative prior as in the keyboard design.  
When the kernel is the Kronecker-delta kernel, i.e., $k(d_j,d_s)=I(j=s)$ indicating no borrowing from neighbouring doses, $w(d_j;d_s)=I(j=s)$ and \eqref{eq:BKPmodel} reduces to \eqref{eq:BetaPost}.
Thus, the keyboard design is recovered as a special case of SKBD.

Unlike model-based approaches such as the continual reassessment method \citep{OQuigley1990CRM} or the Bayesian logistic regression model \citep{Neuenschwander2008BLRM}, SKBD does not impose a parametric model for the global dose-toxicity curve. Instead, it constructs dose-wise posteriors using kernel-weighted borrowing.
As discussed in \citet{Zhao2025BKP}, \eqref{eq:BKPmodel} is induced by a kernel-weighted likelihood formulation and yields a principled, data-adaptive borrowing of information from neighbouring doses.
This perspective fits well in phase~I design, where the primary objective is safe identification of an MTD under limited sample size and without strong justification for a specific global parametric dose--response model.

A central appeal of rule-based designs is their operational simplicity: interim decisions can be reduced to counting DLTs and consulting pre-tabulated boundaries. Although SKBD incorporates information borrowing across doses, it remains a rule-based design rather than a fully model-driven procedure. At any interim, the decision at dose $d_j$ is determined by the BKP posterior in \eqref{eq:BKPmodel}, which depends on the observed data only through the kernel-weighted aggregates $(y'_j,n'_j)$. 
Therefore, with a fixed dose grid and a prespecified kernel, SKBD can be implemented either by computing these aggregates at the time of decision making or by referring to a decision table indexed by $(y'_j,n'_j)$ for a given configuration of interim data at other doses; see Table~\ref{tab:keyboard-vs-skbd}. 
In this way, SKBD retains the transparency and ease of implementation as rule-based design while allowing controlled information sharing across doses.

\subsubsection{Kernel}\label{sec:Kernel}

The kernel specifies how toxicity information is borrowed across dose levels. Intuitively, the degree of borrowing decreases in distance. 

First, we linearly map the prespecified doses onto $[0,1]$ by setting $d_j \leftarrow (d_j-d_1)/(d_J-d_1)$.  (When doses are defined on a fold-change scale, we apply a log transformation before this linear mapping.) 
This preprocessing preserves dose ordering and places doses on a common scale. 

A key safety consideration in phase~I oncology is that \emph{overdosing is typically more harmful than underdosing} \citep{Yuan2016BOIN, Zhang2022BLRM}. 
To reflect this point, SKBD uses an asymmetric borrowing scheme under which toxicity outcomes observed at higher doses may have a stronger effect on inference at a lower dose than outcomes observed in the reverse direction.
To encode this asymmetry, we use an \emph{asymmetric} Gaussian kernel.
For $d,d'\in[0,1]$, define
\begin{equation}\label{eq:asymKernel}
	k_{\rm asym}(d, d') = \begin{cases}
		\exp\{-\theta_1 (d - d')^2\}, & \text{if } d'\leq d,\\
		\exp\{-\theta_2 (d - d')^2\}, & \text{if } d'> d,
	\end{cases}	
\end{equation}
where positive coefficients $\theta_1$ and $\theta_2$ control the rate at which borrowing decays with distance in the two directions.
Because a larger $\theta$ implies faster decay, assigning more weight to higher doses than to lower doses at the same distance is achieved by setting $0<\theta_2<\theta_1$.
When $\theta_1=\theta_2=\theta$, \eqref{eq:asymKernel} reduces to the symmetric Gaussian kernel $k_{\rm sym}(d,d')=\exp\{-\theta(d-d')^2\}$.
Sensitivity analyses comparing the asymmetric kernel with symmetric alternatives are provided in Section~\ref{sec:asyVSsy} of Supplementary Material.

\begin{figure}[!t]
	\centering
	\includegraphics[width=0.8\linewidth]{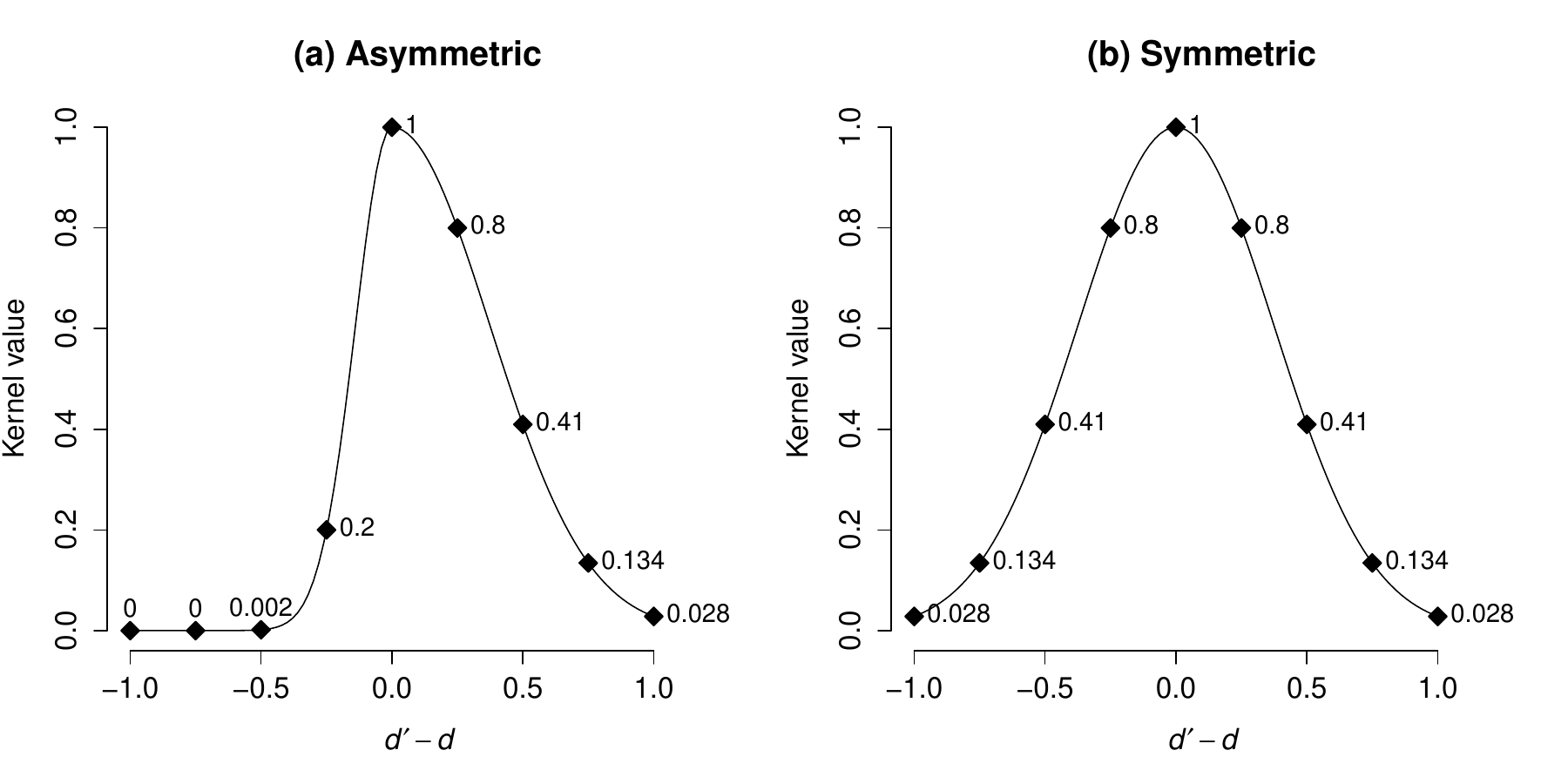}
	\caption{Illustration of kernel functions as a function of $d'-d$: 
		(a) asymmetric kernel $k_{\rm asym}(d,d')$ in (\ref{eq:asymKernel}) 
		with $\theta_1=25.75$ and $\theta_2=3.57$; 
		(b) symmetric kernel $k_{\rm sym}(d,d')$ with $\theta=3.57$.}
	\label{fig:kernel}
\end{figure}

Instead of subjectively specifying $(\theta_1,\theta_2)$, we calibrate them through an interpretable nearest-neighbour quantity. 
After rescaling doses to $[0,1]$, let $\sigma=\min_{j\ge 2}|d_j-d_{j-1}|$ 
be the smallest gap between neighbouring doses. 
(In the equally spaced setting with $d_j=(j-1)/(J-1)$, $\sigma=(J-1)^{-1}$.) For any interior dose $d_j$, the quantities $k_{\rm asym}(d_j,d_j+\sigma)$ and $k_{\rm asym}(d_j,d_j-\sigma)$ represent the pre-normalisation borrowing weights from the higher- and lower-dose directions, respectively. 
As a default specification, we set
\[
k_{\rm asym}(d_j,d_j+\sigma)=0.8, \qquad k_{\rm asym}(d_j,d_j-\sigma)=0.2,
\]
so that the nearest higher dose exerts stronger influence than the nearest lower dose. Under \eqref{eq:asymKernel}, this corresponds to $\theta_2=-\log(0.8)/\sigma^2$ and $\theta_1=-\log(0.2)/\sigma^2$.
Sensitivity analyses examining different nearest-neighbour borrowing specifications are provided in  Section~\ref{sec:senKernel} of Supplementary Material.
Because the Gaussian kernel decays quadratically with distance, the degree of borrowing drops off rapidly beyond adjacent doses. 
For example, in the equally spaced setting, the (pre-normalisation) kernel value for a dose $m$ levels away is $\{k_{\rm asym}(d_j,d_j\pm\sigma)\}^{m^2}$, which amounts to $0.2^4=0.0016$ for a dose that is two levels below the current dose. This property keeps the information sharing predominantly local. 
When dose levels are equally spaced or represented by their ranks, the kernel values are determined by both step distance and direction, as illustrated in Figure~\ref{fig:kernel}.

\subsubsection{Design}\label{sec:design}

\begin{figure}[!t]
	\centering
	\includegraphics[width=\linewidth]{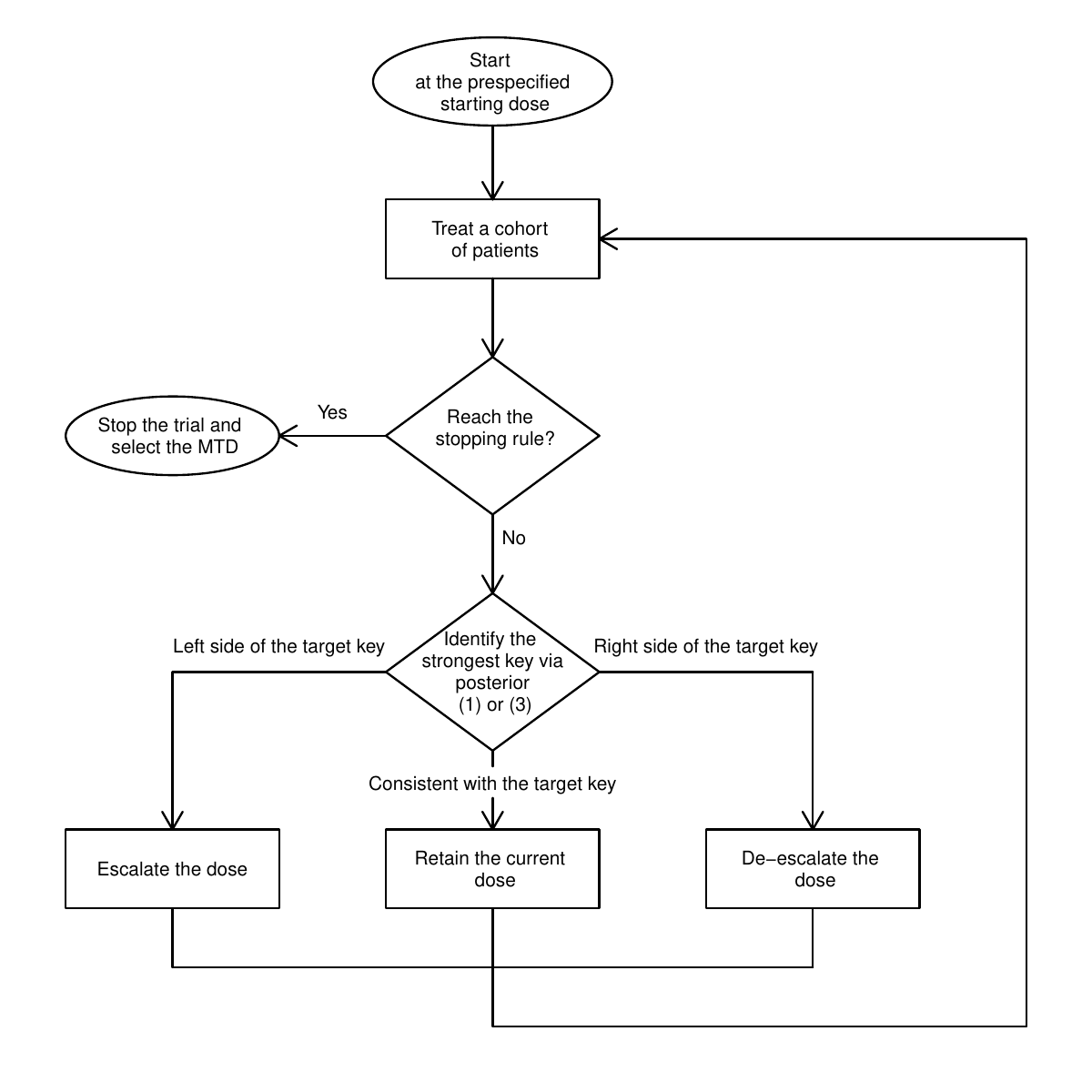}
	\caption{Work flow of the keyboard design and its shared extension (SKBD). Both designs follow the same decision structure and differ only in how the posterior toxicity distribution is constructed.}
	\label{fig:flowchart}
\end{figure}

The shared keyboard design (SKBD) retains the same key-based decision framework as the keyboard design described in Section~\ref{sec:KBD}. The main modification is in the construction of the posterior toxicity distribution: SKBD uses the BKP posterior in \eqref{eq:BKPmodel} instead of the independent beta-binomial posterior in \eqref{eq:BetaPost}. 
The procedure is summarised in Figure~\ref{fig:flowchart}.

At the end of the trial, the final MTD is selected using dose-wise posterior mean toxicity estimates, followed by isotonic regression as in the keyboard design.
To maintain comparability with the final selection convention used in the \texttt{Keyboard} and \texttt{BOIN} packages \citep{Li2022keyboard, Yan2020BOIN}, we compute these posterior mean estimates under a weak $\mathrm{Beta}(0.01,0.01)$ prior and use a symmetric kernel in this step to avoid introducing directional bias into the final dose-wise toxicity estimates. 
Specifically, we use $k_{\rm sym}(d_j,d_j\pm\sigma)=0.2$. 


%

\subsubsection{A Protocol-Style Trial-Planning Illustration}\label{sec:example}

We illustrate how SKBD can be used at the protocol-planning stage of a single-agent phase~I oncology trial. Consider a trial with five prespecified dose levels, cohort size three, and target DLT rate $\phi=0.3$. The practical question is whether toxicity information emerging at neighbouring doses should modify the usual Keyboard recommendation while preserving a transparent, pre-tabulated decision rule.

For this illustration, we consider a representative five-dose configuration with true DLT probabilities $(0.01, 0.12, 0.30, 0.41, 0.55)$. Under this scenario, $d_3$ is the true MTD, whereas the next higher dose $d_4$ is already above the target toxicity region. Therefore, DLTs observed at $d_4$ provide clinically relevant safety information when deciding whether to continue escalation from $d_3$.

\begin{figure}[!t]
	\centering
	\includegraphics[width=0.8\linewidth]{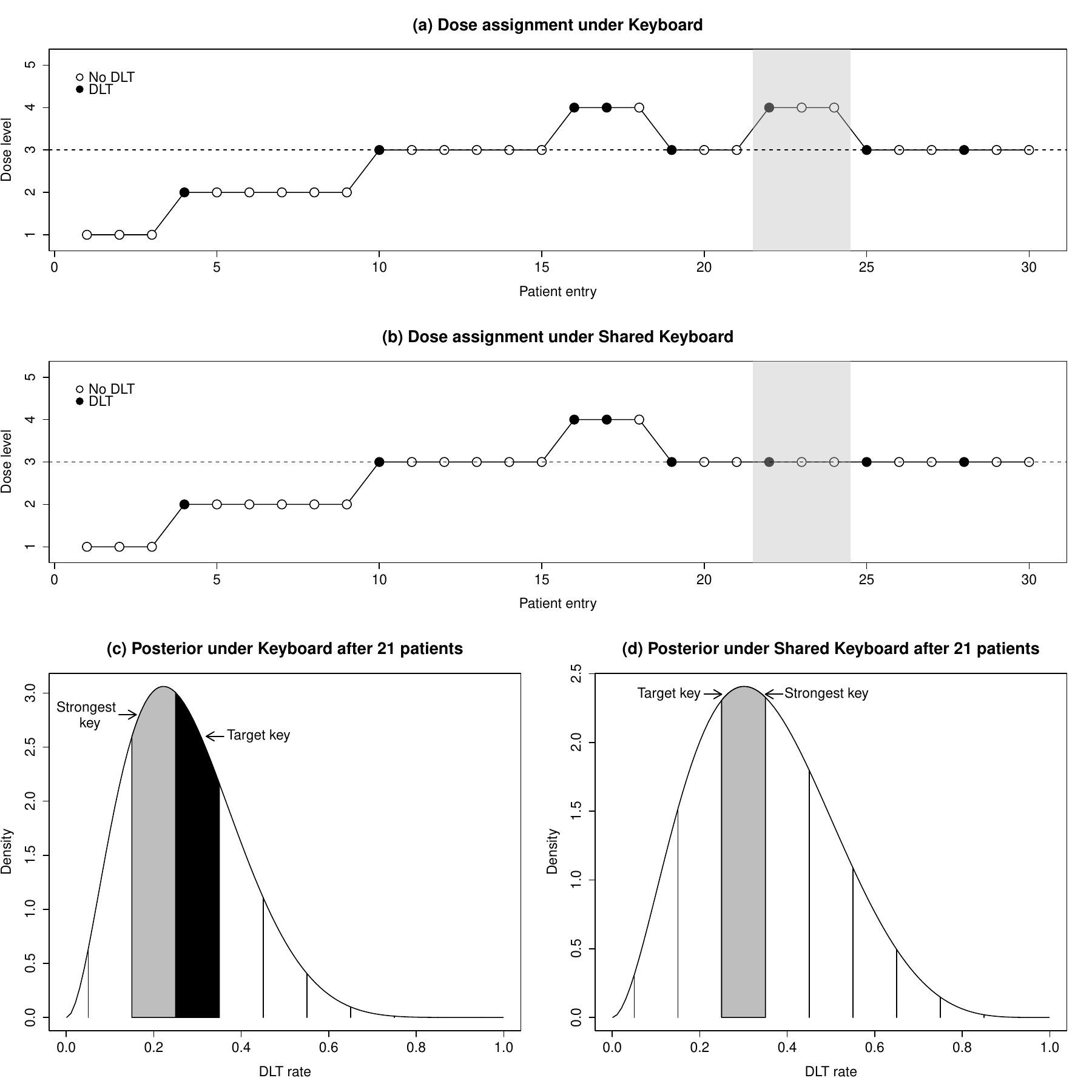} 
	\caption{Dose-assignment paths and interim posterior toxicity distributions under SKBD and the keyboard design in a representative trial (30 patients). Panels (a)--(b): cohort-by-cohort dose assignments (open/filled circles: no DLT/DLT); grey shading highlights the portions where the two designs make different dose-assignment decisions. Panels (c)--(d): posterior distribution of the toxicity probability at the interim decision dose after 21 patients; the black and grey shaded regions indicate the target key and the strongest key, respectively.}
	\label{fig:path}
\end{figure}

Figure \ref{fig:path} compares SKBD with the keyboard design in one representative trial realisation.
Panels (a) and (b) show that the two designs make the same escalation decision until cohort 7   after two DLT events are observed at the adjacent higher dose in cohort 6.
Unlike the keyboard design, SKBD incorporates this information through kernel-weighted borrowing, producing a posterior distribution under which further escalation is prevented as desired.

Panels (c) and (d) explain this difference in detail. After cohort seven at $d_3$, the observed data across levels 1--5 in terms of $(y,n)$ are $(0,3)$, $(1,6)$, $(2,9)$, $(2,3)$, and $(0,0)$, respectively. Under the keyboard design, inference for $\pi_3$ is based only on $(y_3,n_3)=(2,9)$.
Under SKBD, inference for $\pi_3$ is based on $(y_3',n_3')=(1.9,6.3)$ through \eqref{eq:BKPmodel}. By the default asymmetric kernel, the borrowed contribution from dose 4, the nearest higher dose, is larger than that from dose 2, the nearest lower dose.
As a result, SKBD produces a strongest key that coincides with the target key, whereas the keyboard design produces a strongest key that lies to the left of the target key.

\begin{table}[!t]
	\centering
	\caption{Pre-tabulated decision boundaries: keyboard vs.\ SKBD (conditional on interim data at other doses).}
	\label{tab:keyboard-vs-skbd}
	\renewcommand{\arraystretch}{1.15}
	\setlength{\tabcolsep}{3.2pt}
		\begin{tabular}{lcccccccc>{\columncolor{shade}}c ccccccccc}
			\toprule
			\multicolumn{19}{l}{(a) Keyboard: $\phi=0.3$, target key $(0.25,0.35)$} \\
			\midrule
			Number treated & 1&2&3&4&5&6&7&8&9&10&11&12&13&14&15&16&17&18 \\
			Escalate if \#DLT $\le$ & 0&0&0&0&1&1&1&1&2&2&2&2&3&3&3&3&4&4 \\
			De-escalate if \#DLT $\ge$ & 1&1&2&2&2&3&3&3&4&4&4&5&5&5&6&6&6&7 \\
			Eliminate if \#DLT $\ge$ & NA&NA&3&3&4&4&5&5&5&6&6&7&7&8&8&8&9&9 \\
			\midrule
			\multicolumn{19}{l}{(b) SKBD: conditional table at $d=3$, $\phi=0.3$} \\
			\multicolumn{19}{l}{\quad Other-dose data at the interim: $(y,n)=(0,3),(1,6),(2,3),(0,0)$ at doses $1,2,4,5$} \\
			\midrule
			Number treated at dose 3 & 1&2&3&4&5&6&7&8&9&10&11&12&13&14&15&16&17&18 \\
			Escalate if \#DLT $\le$ & NA&NA&NA&0&0&0&0&1&1&1&1&2&2&2&2&3&3&3 \\
			De-escalate if \#DLT $\ge$ & NA&1&1&1&2&2&2&3&3&3&4&4&4&5&5&6&6&6 \\
			Eliminate if \#DLT $\ge$ & NA&NA&NA&4&4&5&5&6&6&7&7&8&8&8&9&9&10&10 \\
			\bottomrule
		\end{tabular}
	{\footnotesize\raggedright \emph{Note}: NA indicates that the corresponding action does not occur at that sample size under the given decision rule. The shaded column highlights $n=9$, corresponding to the interim shown in Figure~\ref{fig:path}. \par}
\end{table}

Table~\ref{tab:keyboard-vs-skbd} complements Figure~\ref{fig:path} by comparing the standard keyboard boundaries \citep[c.f. Panel~(a);][]{Yan2017Keyboard} with the SKBD boundaries at dose~3 conditional on the interim outcomes observed at the other doses in this example (Panel~(b)).
This conditional tabulation shows explicitly how neighbouring-dose information changes the decision boundaries while preserving the transparency of the keyboard framework.
Because dose~4 has already shown substantial toxicity (2 DLTs among 3 patients), SKBD becomes more conservative at dose~3 than the keyboard design.
For example, at $n_3=9$ (the shaded column), SKBD escalates only when $y_3\le 1$, whereas the keyboard design escalates when $y_3\le 2$.
Moreover, under this interim configuration, SKBD has no escalation region when $n_3\le 3$.
These boundary changes provide a transparent explanation for the divergence of the two allocation paths in Figure~\ref{fig:path}.

In Section~\ref{sec:simulation}, we evaluate the design systematically across a broad range of dose--toxicity scenarios and assess how controlled information sharing affects both safety and accuracy.

\subsection{Extension to Dose Insertion}\label{sec:insertion}

\begin{figure}[!t]
	\centering
	\includegraphics[width=0.6\linewidth]{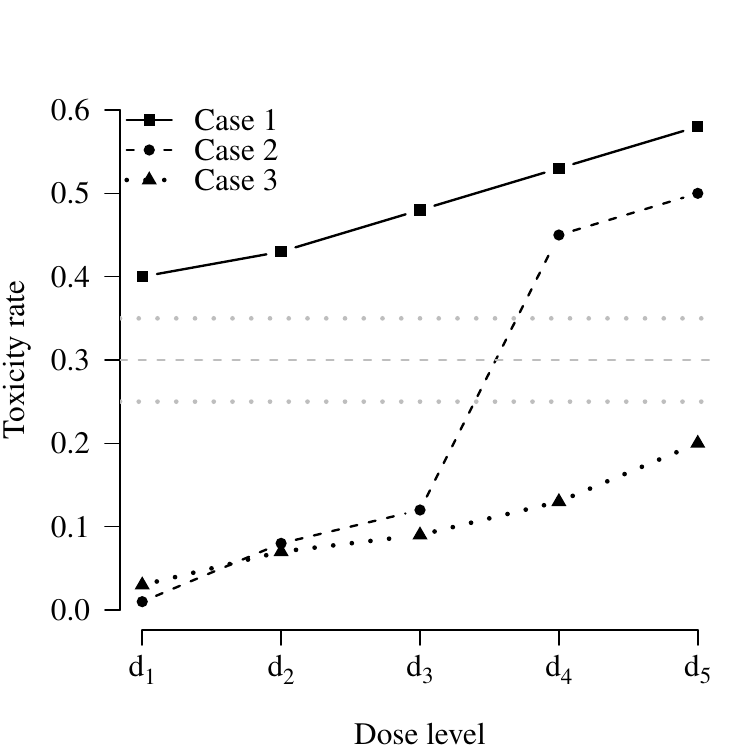}
	\caption{Dose--toxicity curves illustrating situations that may warrant dose insertion. The horizontal dashed line represents the target toxicity rate $\phi=0.3$. The two horizontal dotted lines indicate the target key bounds, $\phi-\varepsilon_1=0.25$ and $\phi+\varepsilon_2=0.35$. }
	\label{fig:demo_insert_curve}
\end{figure}

When accumulating evidence suggests that the true MTD is not well represented by the prespecified doses, it is desirable to refine the dose grid by adaptively inserting a new dose \citep{Hu2013ADI, Chu2016ADM}. 
As illustrated in Figure~\ref{fig:demo_insert_curve}, this need typically arises in three settings: (i) all investigated doses appear overly toxic (case~1); (ii) the target toxicity region lies between two adjacent prespecified doses (case~2); and (iii) all investigated doses appear subtherapeutic (case~3). 
Existing methods address this problem in different ways. 
\citet{Hu2013ADI} proposed a model-based insertion procedure that combines an activation rule with inverse dose--response estimation. 
\citet{Chu2016ADM} proposed to trigger dose insertion using beta--binomial posterior probabilities and to estimate the inserted dose using a nonparametric local polynomial model. 
By extending SKBD, we propose a method, called insertion SKBD (Ins-SKBD), to take a different route. It evaluates candidate insertion locations directly through the same BKP-based borrowing mechanism used for dose transition. 
In this way, dose insertion is incorporated into the SKBD framework without introducing a separate global dose--toxicity model for candidate-dose estimation.

We use the same dose preprocessing procedure as in SKBD for dose insertion. Denote the doses before standardization by $\tilde d_1<\cdots<\tilde d_J$ for late use. 
Interior insertions therefore lie between adjacent doses, whereas boundary insertions may yield standardised locations below $0$ or above $1$.

Let $\mathcal S=\{s:n_s>0\}$ denote the set of dose levels at which at least one patient has been treated. 
To determine whether dose insertion is warranted at location $d$, we construct a Beta distribution based on kernel-weighted pseudo-counts as 
\begin{equation}\label{eq:BKP_insert_check}
	\pi_{\mathrm{ins}}(d)\mid\mathcal D \sim
	\mathrm{Beta}\!\left(
	a(d)+\sum_{s\in\mathcal S} w_{\mathrm{ins}}(d;d_s)\,y_s,\;
	b(d)+\sum_{s\in\mathcal S} w_{\mathrm{ins}}(d;d_s)\,(n_s-y_s)
	\right),
\end{equation}
where $w_{\mathrm{ins}}(d;d_s)$ denotes normalised kernel weights induced by a symmetric kernel.
We set $a(d)=b(d)=0.5$ for all $d$ following \citet{Chu2016ADM}. 

 
At each interim analysis, dose insertion is considered before dose escalation (by SKBD).
Let $\mathcal G=\{d_{(1)}<\cdots<d_{(m)}\}$ denote the current working dose set, consisting of the prespecified doses together with any doses inserted earlier in the trial, and let $d_j$ denote the current dose. 
Based on \eqref{eq:BKP_insert_check}, dose insertion is triggered if one of the following conditions holds: (i) if $d_j=d_{(1)}$,
\[
	\prob\{\pi_{\mathrm{ins}}(d_{(1)})>\phi+\varepsilon_2\mid\mathcal D\}>C_2;
\]
(ii) for some $r=1,\ldots,m-1$,
\[
	\prob\{\pi_{\mathrm{ins}}(d_{(r)})<\phi-\varepsilon_1\mid\mathcal D\}>C_1
\quad\text{and}\quad
\prob\{\pi_{\mathrm{ins}}(d_{(r+1)})>\phi+\varepsilon_2\mid\mathcal D\}>C_2;
\]
(By \citet{Chu2016ADM}, condition (ii) can identify at most one adjacent interval when $C_1+C_2>1$.) 
and (iii) if $d_j=d_{(m)}$,
\[
	\prob\{\pi_{\mathrm{ins}}(d_{(m)})<\phi-\varepsilon_1\mid\mathcal D\}>C_1.
\]
These correspond to lower-boundary, interior, and upper-boundary insertion, respectively, where $C_1$ and $C_2$ are prespecified probability cutoffs. As in \citet{Chu2016ADM}, we set $C_1=C_2=0.6$ throughout. To respect monotonicity, the relevant posterior probabilities are isotonic-adjusted over dose before these criteria are evaluated.

If an interior insertion is triggered between $d_{(r)}$ and $d_{(r+1)}$, let
\[
	q(d)=\prob\{\phi-\varepsilon_1<\pi_{\mathrm{ins}}(d)\le \phi+\varepsilon_2\mid\mathcal D\},
\]
denote the posterior probability that dose $d$ lies in the target key. We then select the inserted dose as
\begin{equation}
	d^{\mathrm{new}} = \underset{d\in(d_{(r)},\,d_{(r+1)})}{\argmax}\, q(d).
\end{equation}
For boundary insertions, we determine the inserted dose on the original dose scale as
\[
\tilde d^{\mathrm{new}}=
\begin{cases}
	\tilde d_{(1)}/2, & \text{for lower-boundary insertion},\\[1mm]
	3\tilde d_J/2, & \text{for upper-boundary insertion},
\end{cases}
\]
where $\tilde d_{(1)}$ is the current minimum dose (after possible insertion in previous steps) and $\tilde d_J$ is the highest prespecified dose on the original scale. 
The use of $\tilde d_J$ rather than the current maximum dose in the upper-boundary case prevents repeated multiplicative extrapolation beyond the prespecified dose range.
The inserted dose is then mapped to the standardised scale and incorporated into the working dose grid for subsequent analyses. 

After insertion, the working grid is augmented and re-indexed, and the next cohort is treated at the new dose. If no insertion is triggered, the trial proceeds according to the ordinary SKBD decision rule. If the current dose is declared overly toxic and is the lowest dose on the working grid, the trial stops early for safety; dose insertion does not override this stopping rule.

\subsection{Extension to Late-Onset Toxicity}\label{sec:TITE-SKBD}

In phase I trials of molecularly targeted agents or immunotherapies, toxicities may be late onset relative to the patient accrual rate, so that some DLT outcomes remain pending at interim decision times \citep{PostelVinay2011LateOnset, Weber2015LateOnset}. 
To handle this issue without suspending accrual, we extend SKBD to late-onset toxicity settings using a time-to-event (TITE) framework \citep{Yuan2018TITEBOIN, Zhou2024TITE}.

Specifically, we adopt the effective sample size (ESS) approach of \citet{Lin2020ESS} to convert partially observed follow-up to effective dose-level toxicity data. 
These effective data replace the complete-data binomial counts in SKBD, while the information-sharing step across doses and the strongest-key decision rule remain unchanged. 
Thus, TITE-SKBD maintains the decision structure of SKBD while permitting interim decisions in the presence of pending DLT outcomes. 

For the sake of space, the formulation, implementation details, and simulation results for TITE-SKBD are provided in Supplementary Material~\ref{sec:SM-TITE-SKBD}.

\subsection{Software}

The R code for implementing the proposed methods is provided in the R package \texttt{SKBD}, which is publicly available on GitHub  (\url{https://github.com/Jiangyan-Zhao/SKBD}). 
The package allows users to generate SKBD decision tables, conduct simulation studies, obtain operating characteristics, and implement the dose-insertion and time-to-event extensions described in this article.
	
To facilitate the use of SKBD in practical phase I trial design, we further developed a Shiny-based graphical user interface. 
The Shiny application enables users to easily specify key design parameters, control the shared-borrowing mechanism, generate decision tables, run operating-characteristic simulations, and visualise dose-assignment and dose-insertion results. 
Screenshots of the software interface are provided in Figures~\ref{fig:shiny-setting} and \ref{fig:shiny-simulation} of the Supplementary Material.

\section{Simulation studies}\label{sec:simulation}

We evaluated the operating characteristics of the proposed designs through simulation studies.
Unless otherwise stated, patients were enrolled in cohorts of three, with a maximum sample size of 30 (10 cohorts), and all results were based on 10{,}000 replicated trials per scenario.
We first assessed SKBD under a fixed prespecified dose grid, which provided the primary setting for evaluating the effect of kernel-based information sharing on dose-escalation decisions. 
We then considered the dose-insertion extension, Ins-SKBD, in settings where the initial dose grid did not adequately cover the target toxicity region and adaptive grid refinement was warranted.
The simulation results for the late-onset toxicity extension, TITE-SKBD, are provided in Supplementary Material, Section~\ref{sec:OC-TITE}.

\subsection{Performance of SKBD}

\subsubsection{Setup}

We considered two sets of scenarios to evaluate the operating characteristics of SKBD: i) 20 fixed toxicity scenarios and ii) 1{,}000 randomly generated dose--toxicity curves.  

\begin{enumerate}[label=(\roman*), itemsep=2pt, topsep=0pt, parsep=0pt]
	\item \textit{Fixed scenarios.}
	We adopted the 20 fixed toxicity scenarios of \citet{Yan2017Keyboard}.
	Each scenario involved five prespecified dose levels and represented a clinically plausible dose--toxicity relationship.
	The target toxicity rate was $0.2$ for scenarios~1--10 and $0.3$ for scenarios~11--20. Across these scenarios, the true MTD ranged from dose 1 to dose 5.
	
	\item \textit{Random scenarios.}
	We further examined performance under 1{,}000 dose--toxicity curves generated using the pseudo-uniform algorithm of \citet{Clertant2017SPM}.
	To ensure a unique and clinically meaningful MTD, we required the true DLT probability at the MTD to lie within the target key $(\phi-0.05,\phi+0.05)$, and the toxicity increments between the MTD and its adjacent doses to be no smaller than 0.05 and no greater than 0.3 \citep{Zhou2021iBOIN}.
\end{enumerate}

Additional details on scenario generation are provided in Supplementary Material, Section~\ref{sec:scenarios}.

\subsubsection{Operating characteristics}

We summarise the performance of the designs using two accuracy measures and two safety measures.

\begin{enumerate}[label=(\roman*), itemsep=2pt, topsep=0pt, parsep=0pt]
	\item \emph{Percentage of correct selection (PCS):}
	the proportion of simulated trials in which the selected dose was the true MTD; trials that terminated early without selecting an MTD were counted as incorrect selections.
	
	\item \emph{Percentage of correct allocation (PCA):}
	the mean proportion of enrolled patients treated at the true MTD, computed within each trial as the number of patients treated at the true MTD divided by the realised sample size. 
	
	\item \emph{Percentage treated above the MTD (above-MTD):}
	the mean proportion of patients treated at doses above the true MTD out of the treated patients.
	
	\item \emph{Risk of overdosing (ROD):}
	the proportion of simulated trials in which at least 60\% of patients were treated above the true MTD.
\end{enumerate}

\subsubsection{Results}

\paragraph{Fixed scenarios.}

\begin{figure}[!t]
	\centering
	\includegraphics[width=0.8\linewidth]{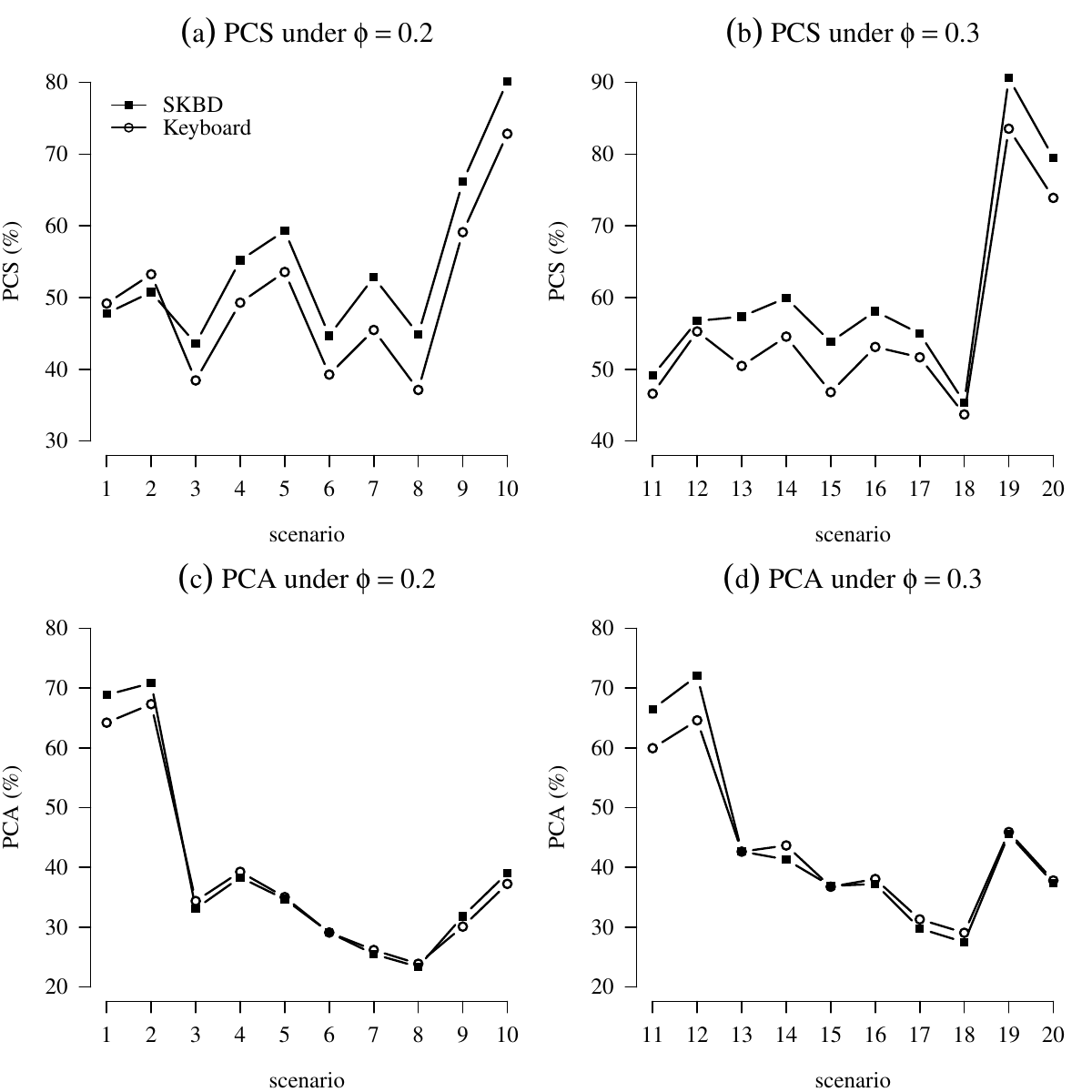}
	\caption{Accuracy performance under fixed scenarios. Panels (a) and (b) show PCS for $\phi=0.2$ and $\phi=0.3$, respectively; panels (c) and (d) show PCA. A higher value is better.}
	\label{fig:fixed-accuracy}
\end{figure}

\begin{figure}[!t]
	\centering
	\includegraphics[width=0.8\linewidth]{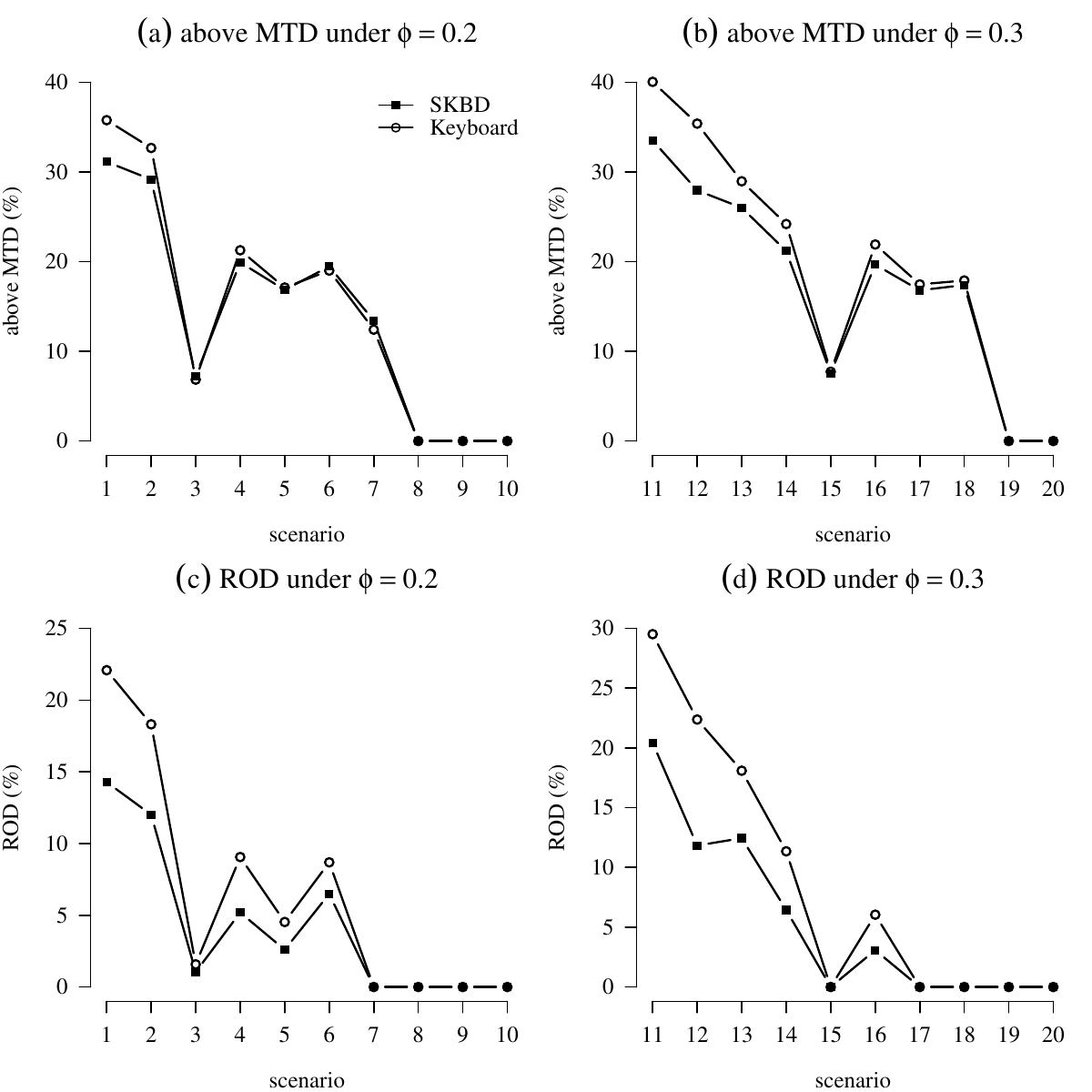}
	\caption{Safety performance under fixed scenarios. Panels (a) and (b) show above-MTD for $\phi=0.2$ and $\phi=0.3$, respectively; panels (c) and (d) show ROD. A lower value is better.}
	\label{fig:fixed-safety}
\end{figure}

Figures~\ref{fig:fixed-accuracy} and~\ref{fig:fixed-safety} summarise the operating characteristics of SKBD and the keyboard design across the 20 fixed toxicity scenarios. The details for each scenario are reported in Tables~\ref{tab:fixed-accuracy} and~\ref{tab:fixed-safety}.

\emph{Accuracy.}
Across the fixed scenarios, SKBD improved MTD identification over the keyboard design.
As shown in Figure~\ref{fig:fixed-accuracy} and Table~\ref{tab:fixed-accuracy}, the average PCS increased from 49.8\% to 54.5\% when $\phi=0.2$ and from 56.0\% to 60.5\% when $\phi=0.3$.
The improvement was evident in most scenarios and was particularly pronounced in the scenarios where the target dose was not at the boundary. 

A similar, though more modest, pattern was observed for patient allocation.
The average PCA increased from 38.7\% to 39.5\% in the scenarios with $\phi=0.2$ and from 43.0\% to 43.7\% in the scenarios with $\phi=0.3$.

\emph{Safety.}
Figure~\ref{fig:fixed-safety} and Table~\ref{tab:fixed-safety} show that SKBD also had a more favourable safety profile.
For the low-target scenarios ($\phi=0.2$), the average percentage of patients treated above the MTD decreased from 14.5\% under the keyboard design to 13.7\% under SKBD, and the average ROD decreased from 6.4\% to 4.2\%.
For the high-target scenarios ($\phi=0.3$), the corresponding reductions were from 19.4\% to 17.0\% for above-MTD and from 8.7\% to 5.4\% for ROD.
The reduction in ROD was especially notable in Scenarios 1, 2, and 11--14, where the ROD under the keyboard design exceeded 10\%.
This corroborates that the asymmetric borrowing scheme facilitates moderate escalation when neighbouring higher doses exhibit excessive toxicity.

\emph{Monotonicity.}
SKBD also yielded posterior mean estimates of DLT probability that more frequently respected the monotone dose--toxicity assumption prior to isotonic adjustment.
In particular, the proportion of trials with monotone dose-wise posterior mean estimates increased from 62.8\% to 73.4\% when $\phi=0.2$ and from 70.2\% to 78.5\% when $\phi=0.3$.
This pattern is consistent with the intended stabilising effect of kernel-based information sharing on dose-level inference.

We repeated the simulation using single-patient cohorts and found similar results. The details are provided in Section~\ref{sec:cohort1} of Supplementary Material. These results indicate that SKBD works well across different cohort sizes.

\paragraph{Random scenarios.}

\begin{figure}[!t]
	\centering
	\includegraphics[width=0.8\linewidth]{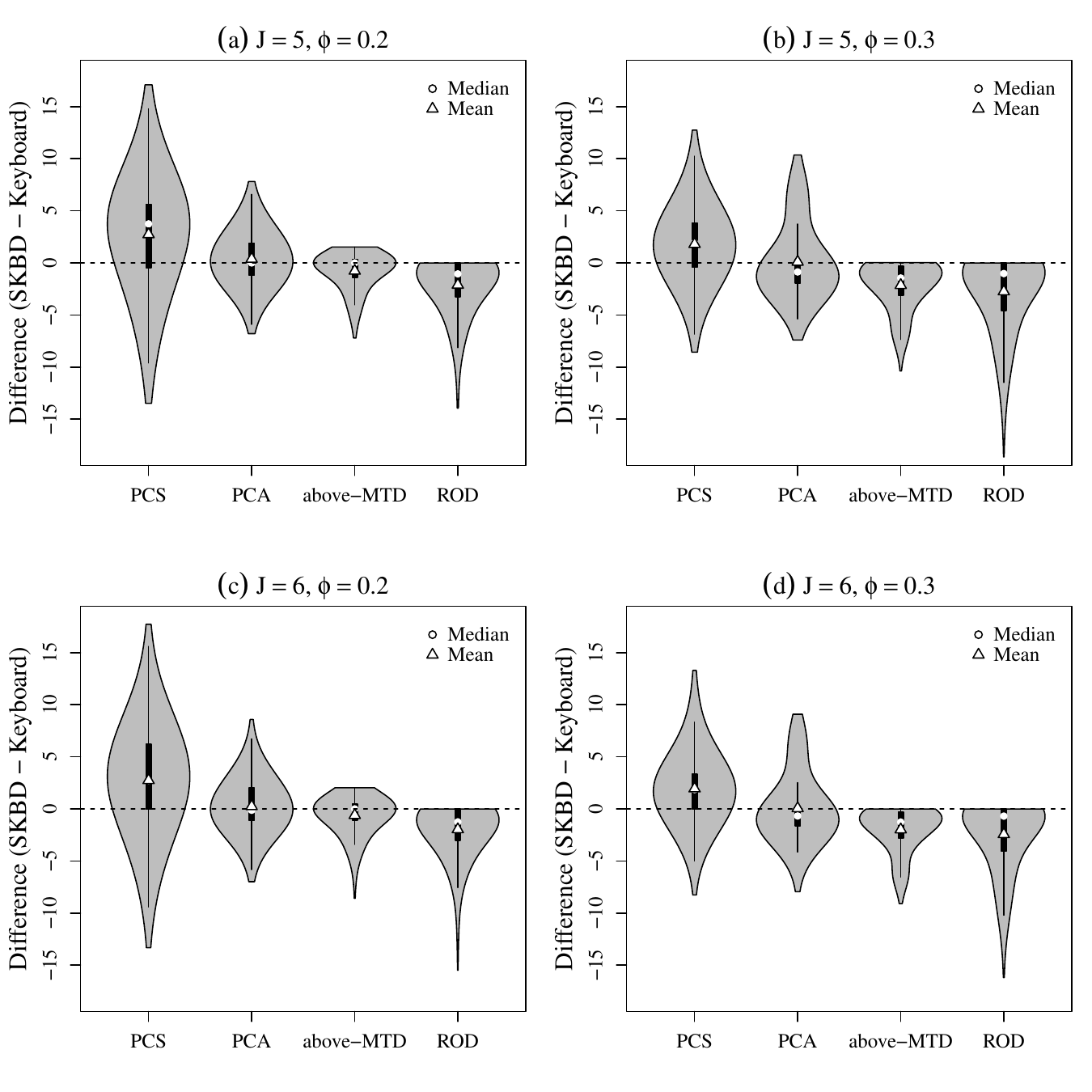}
	\caption{Differences in operating characteristics (SKBD$-$Keyboard), measured in percentage points, under 1,000 randomly generated monotone dose--toxicity scenarios. Panels correspond to different combinations of the number of dose levels ($J$) and target toxicity rate ($\phi$). Violin plots show the across-scenario distributions; circles denote medians and triangles denote means. Positive values favour SKBD for PCS and PCA, whereas negative values favour SKBD for above-MTD and ROD.}
	\label{fig:random}
\end{figure}

The average PCS values of SKBD across 1{,}000 randomly generated monotone dose--toxicity scenarios, stratified by the number of dose levels ($J$) and the target toxicity rate ($\phi$), were 52.1\%, 54.0\%, 48.8\%, and 50.7\% for $(J,\phi)=(5,0.2)$, $(5,0.3)$, $(6,0.2)$, and $(6,0.3)$, respectively. Figure~\ref{fig:random} summarises the differences in all operating characteristics between SKBD and the keyboard design. 
The results were broadly consistent with those obtained under the fixed scenarios. Across all four strata, the distributions of the differences (in percent) were mostly greater than zero for PCS and centred around zero for PCA. By contrast, the distributions of the differences (in percent) for above-MTD and ROD were generally shifted below zero. 
All these findings showed that SKBD yielded a stable improvement in MTD identification across a broad class of scenarios while reducing the risk of overdose.

\subsection{Performance of Ins-SKBD}

\subsubsection{Setup}

We set the target toxicity rate at $\phi=0.3$. We considered six dose--toxicity scenarios in the first rows of Tables~\ref{tab:insert-interior} and~\ref{tab:insert-boundary} where the initial prespecified doses did not adequately bracket the target toxicity region. For the four scenarios in Table~\ref{tab:insert-interior}, the insertion took place between some given doses. For the two scenarios in Table~\ref{tab:insert-boundary}, the insertion took place beyond the boundary doses.
All scenarios were generated from the sigmoidal Emax model \citep{Macdougall2006Emax},
\[
\pi(d)=E_0+\frac{E_{\max}\cdot d^\gamma}{EC_{50}^\gamma+d^\gamma},
\]
with $E_0=0$ and $E_{\max}=1$, and varying values of $(EC_{50},\gamma)$. 

We compared the proposed Ins-SKBD design with the adaptive dose modification (ADM) method of \citet{Chu2016ADM}.  

\subsubsection{Performance metrics}


We evaluated the performance of dose insertion using the following four measures:

\begin{enumerate}[label=(\roman*), itemsep=2pt, topsep=0pt, parsep=0pt]
	\item \emph{Modification rate}: the proportion of simulated trials in which a new dose was inserted.
	
	\item \emph{Inserted-dose location}: the mean and standard deviation (SD) of the inserted dose among trials in which insertion occurred.
	
	\item \emph{Inserted-dose selection rate}: the proportion of simulated trials in which the inserted dose was ultimately selected as the MTD.
	
	\item \emph{Inserted-dose allocation rate}: the mean proportion of enrolled patients treated at the inserted dose across simulated trials.
\end{enumerate}

For completeness, we also reported the dose-selection percentages and patient-allocation percentages for the prespecified doses. These measures show how well dose insertion improves accuracy by shifting final selection toward the target region and how much it improves safety by reducing patient allocation to overly toxic doses.

\subsubsection{Results}

\begin{table}[!t]
	\centering
	\renewcommand{\arraystretch}{1.2}
	\caption{Simulation study comparing Ins-SKBD and ADM in interior dose-insertion scenarios (Scenarios 1--4), with a target toxicity rate of $\phi=0.3$.}
	\label{tab:insert-interior}
	\resizebox{\textwidth}{!}{%
		\begin{tabular}{l cccccc ccc}
			\toprule
			Method & \multicolumn{6}{c}{Prespecified dose} & \multicolumn{3}{c}{Added dose} \\
			\cmidrule(lr){2-7}\cmidrule(lr){8-10}
			
			\multicolumn{10}{c}{ Scenario 1, MTD = 9.6} \\
			\addlinespace[2pt]
			& Toxicity & 0.14 & 0.45 & 0.63 & 0.74 & 0.80 & Mean & \% Sel & \% Modification \\
			& Dose (mg) & 5.00 & 15.00 & 25.00 & 35.00 & 45.00 & (SD) & (\% Pts) & \\
			Ins-SKBD & \% Sel & 5.04 & 11.86 & 0.87 & 0.02 & 0.00 & 10.72 & 82.21 & 96.45 \\
			& \% pts & 26.02 & 19.74 & 3.47 & 0.19 & 0.00 & (3.62) & (50.58) & \\
			ADM      & \% Sel & 8.06 & 15.50 & 0.82 & 0.02 & 0.00 & 12.00 & 75.60 & 96.33 \\
			& \% pts & 20.89 & 25.33 & 3.94 & 0.19 & 0.00 & (3.51) & (49.65) & \\
			\midrule
			
			\multicolumn{10}{c}{ Scenario 2, MTD = 15.1} \\
			\addlinespace[2pt]
			& Toxicity & 0.03 & 0.14 & 0.45 & 0.75 & 0.91 & Mean & \% Sel & \% Modification \\
			& Dose (mg) & 5.00 & 10.00 & 20.0 & 35.00 & 60.00 & (SD) & (\% Pts) & \\
			Ins-SKBD & \% Sel & 0.37 & 5.74 & 13.26 & 0.35 & 0.00 & 15.86 & 80.28 & 96.38 \\
			& \% pts & 11.47 & 24.16 & 19.11 & 2.86 & 0.04 & (3.84) & (42.37) & \\
			ADM      & \% Sel & 0.01 & 9.70 & 19.63 & 0.28 & 0.00 & 16.78 & 70.38 & 94.73 \\
			& \% pts & 11.08 & 20.53 & 24.77 & 3.17 & 0.04 & (3.94) & (40.41) & \\
			\midrule
			
			\multicolumn{10}{c}{ Scenario 3, MTD = 19.6} \\
			\addlinespace[2pt]
			& Toxicity & 0.03 & 0.06 & 0.20 & 0.50 & 0.80 & Mean & \% Sel & \% Modification \\
			& Dose (mg) & 5.00 & 7.50 & 15.00 & 30.00 & 60.00 & (SD) & (\% Pts) & \\
			Ins-SKBD & \% Sel & 0.25 & 1.04 & 10.98 & 10.08 & 0.17 & 22.06 & 77.48 & 94.04 \\
			& \% pts & 10.84 & 12.72 & 23.45 & 13.74 & 1.66 & (6.71) & (37.59) & \\
			ADM      & \% Sel & 0.01 & 0.24 & 20.38 & 15.88 & 0.17 & 23.10 & 63.32 & 91.81 \\
			& \% pts & 10.89 & 12.43 & 24.10 & 18.69 & 1.84 & (6.09) & (32.06) & \\
			\midrule
			
			\multicolumn{10}{c}{ Scenario 4, MTD = 6.8} \\
			\addlinespace[2pt]
			& Toxicity & 0.02 & 0.03 & 0.09 & 0.20 & 0.45 & Mean & \% Sel & \% Modification \\
			& Dose (mg) & 1.00 & 1.50 & 3.00 & 5.00 & 10.00 & (SD) & (\% Pts) & \\
			Ins-SKBD & \% Sel & 0.08 & 0.30 & 0.88 & 10.14 & 12.94 & 7.64 & 75.66 & 91.08 \\
			& \% pts & 10.48 & 11.05 & 14.05 & 17.63 & 13.69 & (2.79) & (33.10) & \\
			ADM      & \% Sel & 0.00 & 0.01 & 1.69 & 22.31 & 23.13 & 7.94 & 52.86 & 82.05 \\
			& \% pts & 10.46 & 11.10 & 14.31 & 21.60 & 20.08 & (2.17) & (22.46) & \\
			\bottomrule
		\end{tabular}
	}
\end{table}
 
 \begin{table}[!t]
 	\centering
 	\renewcommand{\arraystretch}{1.2}
 	\caption{Simulation study comparing Ins-SKBD and ADM in boundary dose-insertion scenarios (Scenarios 5--6), with a target toxicity rate of $\phi=0.3$.}
 	\label{tab:insert-boundary}
 	\resizebox{\textwidth}{!}{%
 		\begin{tabular}{l cccccc ccc}
 			\toprule
 			Method & \multicolumn{6}{c}{Prespecified dose} & \multicolumn{3}{c}{Added dose} \\
 			\cmidrule(lr){2-7}\cmidrule(lr){8-10}
 			
 			\multicolumn{10}{c}{ Scenario 5 (lower-boundary insertion), MTD = 3.2} \\
 			\addlinespace[2pt]
 			& Toxicity & 0.45 & 0.55 & 0.61 & 0.65 & 0.68 & Mean & \% Sel & \% Modification \\
 			& Dose (mg) & 10.00 & 20.00 & 30.00 & 40.00 & 50.00 & (SD) & (\% Pts) & \\
 			Ins-SKBD & \% Sel & 11.23 & 1.13 & 0.10 & 0.00 & 0.00 & 6.47 & 87.54 & 98.29 \\
 			& \% pts & 24.13 & 4.53 & 0.55 & 0.04 & 0.00 & (3.47) & (70.74) & \\
 			ADM      & \% Sel & 13.75 & 1.42 & 0.17 & 0.01 & 0.00 & 8.12 & 84.65 & 97.99 \\
 			& \% pts & 27.16 & 5.11 & 0.56 & 0.04 & 0.00 & (3.72) & (67.13) & \\
 			\midrule
 			
 			\multicolumn{10}{c}{ Scenario 6 (upper-boundary insertion), MTD = 86.8} \\
 			\addlinespace[2pt]
 			& Toxicity & 0.03 & 0.05 & 0.09 & 0.15 & 0.20 & Mean & \% Sel & \% Modification \\
 			& Dose (mg) & 5.00 & 10.00 & 20.00 & 35.00 & 50.00 & (SD) & (\% Pts) & \\
 			Ins-SKBD & \% Sel & 0.23 & 0.48 & 0.50 & 3.12 & 13.23 & 61.44 & 82.44 & 92.08 \\
 			& \% pts & 10.77 & 11.60 & 13.43 & 15.34 & 15.61 & (17.49) & (33.25) & \\
 			ADM      & \% Sel & 0.01 & 0.16 & 1.77 & 8.68 & 47.61 & 56.73 & 41.77 & 66.49 \\
 			& \% pts & 10.89 & 11.89 & 14.02 & 16.99 & 29.42 & (11.31) & (16.80) & \\
 			\bottomrule
 		\end{tabular}
 	}
 \end{table}

Tables~\ref{tab:insert-interior} and~\ref{tab:insert-boundary} compare the six measures obtained by  Ins-SKBD and ADM under interior and boundary dose-insertion scenarios, respectively. 

In Scenario~1, the true MTD is 9.6 mg, which lies between the first and second prespecified doses. Ins-SKBD and ADM yielded nearly identical insertion rates (96.45\% versus 96.33\%), but Ins-SKBD inserted a dose closer to the true MTD on average (10.72 mg versus 12.00 mg) and selected the inserted dose more frequently as the final MTD (82.21\% versus 75.60\%). 
In addition, Ins-SKBD allocated fewer patients to the overly toxic doses (23.40\% versus 29.46\%) and slightly more patients to the inserted dose (50.58\% versus 49.65\%).
In Scenario~2, where the true MTD is 15.1 mg and lies between the second and third prespecified doses, the average inserted dose under Ins-SKBD was again closer to the true MTD than that under ADM (15.86 mg versus 16.78 mg), with a higher inserted-dose selection rate (80.28\% versus 70.38\%) and a slightly higher insertion rate (96.38\% versus 94.73\%).

Under the remaining interior insertion scenarios (Scenarios~3--4), we observed a similar pattern. 
In Scenario~3, Ins-SKBD inserted a dose closer to the true MTD on average (22.06 mg versus 23.10 mg), selected the inserted dose more frequently as the final MTD (77.48\% versus 63.32\%), and treated more patients at the inserted dose (37.59\% versus 32.06\%), while allocating fewer patients to the two overly toxic doses (15.40\% versus 20.53\%).
The gain was particularly notable in Scenario~4, where the inserted-dose selection rate increased from 52.86\% under ADM to 75.66\% under Ins-SKBD. This improvement was accompanied by a safer patient-allocation pattern. Compared with ADM, Ins-SKBD treated fewer patients at the highest prespecified dose (13.69\% versus 20.08\%), which was above the target region, and more patients at the inserted dose (33.10\% versus 22.46\%).

In Scenario~5 where the true MTD (3.2 mg) lies below the lowest prespecified dose, the two methods had nearly identical insertion rates (98.29\% versus 97.99\%), but Ins-SKBD achieved a slightly higher inserted-dose selection rate (87.54\% versus 84.65\%) and yielded an average inserted dose closer to the true MTD (6.47 mg versus 8.12 mg). Ins-SKBD also allocated a slightly smaller proportion of patients to the overly toxic prespecified doses and a larger proportion to the inserted dose (70.74\% versus 67.13\%). 
In Scenario~6 where the true MTD (86.8 mg) lies above the highest prespecified dose, the advantage of Ins-SKBD was more pronounced. Compared with ADM, Ins-SKBD achieved a substantially higher inserted-dose selection rate (82.44\% versus 41.77\%), allocated considerably more patients to the inserted dose (33.25\% versus 16.80\%), maintained a much higher insertion rate (92.08\% versus 66.49\%), and also yielded an average inserted dose closer to the true MTD (61.44 mg versus 56.73 mg).

In summary, the simulation results showed that Ins-SKBD was superior to ADM in accurately identifying and selecting a desired inserted dose and in overdose control. 
The proposed insertion rule effectively refined an inadequate initial dose grid while retaining the conservative, overdose-aware operating behaviour of SKBD.

\section{Discussion}\label{sec:discussion}

We propose the shared keyboard design for phase I clinical trials.
It generalises the Bayesian model-assisted Keyboard design \citep{Yan2017Keyboard} by replacing single-dose stand-alone posterior inference with a kernel-weighted neighbourhood information-sharing mechanism. This information-sharing strategy enables the SKBD to outperform the Keyboard design in both accuracy and overdose control, while largely preserving its decision structure and operational transparency. 
We further extend the design to handle adaptive dose insertion (Ins-SKBD) and late-onset toxicity (TITE-SKBD) and demonstrate the improvement over their respective counterparts.

A few limitations are noted. For SKBD, the Gaussian kernel may be inferior to others when dose spacing is highly irregular or stronger prior knowledge is available. For Ins-SKBD, the candidate inserted dose is statistically selected and may require feasibility checks, such as formulation constraints and clinical safety review, before implementation. For TITE-SKBD, it relies on the effective sample size approximation and therefore does not explicitly model the event-time distribution. 

Lastly, we note that the idea of kernel-weighted information sharing can be extended to other interval-based designs, such as BOIN \citep{Liu2015BOIN} and i3+3 \citep{Liu2020i3plus3}. Several further extensions also warrant investigation, including designs for drug-combination trials, where the dose space is two- or multi-dimensional \citep{Pan2020ComKeyboard}, and designs for identifying the optimal biological dose, where responses are bivariate \citep{Shi2021uTPI, FDA2022Optimus}. 

\section*{Data availability}
No new data were created or analysed in this study. The R package implementing the proposed methods is publicly available at \url{https://github.com/Jiangyan-Zhao/SKBD}. The R scripts, simulation outputs, figures and tables used to reproduce the results in this article are available from the companion reproducibility repository at \url{https://github.com/Jiangyan-Zhao/SKBD-paper}.

\section*{Supplementary material}
Supplementary material is available online and contains additional analyses of SKBD, including fixed- and random-scenario results, kernel-sensitivity analyses, and cohort-size sensitivity analyses. It also provides methodological details and simulation results for TITE-SKBD, together with screenshots of the Shiny-based software interface.

\bibliographystyle{chicago}
\bibliography{DF}

\newpage
\thispagestyle{empty}

{
	\renewcommand{\thefootnote}{\fnsymbol{footnote}}
	
	\begin{center}
		\ 
		\vskip 2cm
		\Large{Supplementary Materials to ``Shared Keyboard: An improved Bayesian design for phase I clinical trials via Beta kernel process''}
	\end{center}
	
	\begin{center}
		Jiangyan Zhao$^{1}$, Xian Shi$^{1}$, and Jin Xu$^{1,2,}$\footnotemark[1]
	\end{center}
	
	\noindent $^1$School of Statistics, East China Normal University, Shanghai, China\\
	$^2$Key Laboratory of Advanced Theory and Application in Statistics and Data Science -- MOE, East China Normal University, Shanghai, China
	
	\footnotetext[1]{Corresponding author: School of Statistics, East China Normal University, 3663 North Zhongshan Road, Shanghai 200062, China, email: jxu@stat.ecnu.edu.cn}
	
	\begin{abstract}
		Section~\ref{sec:addSKBD} provides additional analyses of SKBD, including the fixed and randomly generated dose--toxicity scenarios, scenario-specific accuracy and safety results for the fixed scenarios, comparisons between asymmetric and symmetric kernels, sensitivity analyses for nearest-neighbour kernel values, and sensitivity to cohort size. Section~\ref{sec:SM-TITE-SKBD} gives the details of the time-to-event extension, TITE-SKBD, including the effective-sample-size construction for pending toxicity outcomes and the corresponding operating characteristics under late-onset toxicity. Section~\ref{sec:software} describes the Shiny-based software interface for implementing SKBD, generating decision tables, and conducting simulation studies.
	\end{abstract}
}

\setcounter{footnote}{0}
\renewcommand{\thefootnote}{\arabic{footnote}}

\newpage
\appendix

\renewcommand{\thesection}{SM\S\arabic{section}}
\renewcommand{\thesubsection}{SM\S\arabic{section}.\arabic{subsection}}
\renewcommand{\thesubsubsection}{SM\S\arabic{section}.\arabic{subsection}.\arabic{subsubsection}}
\setcounter{figure}{0}
\setcounter{table}{0}
\setcounter{equation}{0}
\renewcommand{\thefigure}{SM.\arabic{figure}}
\renewcommand{\thetable}{SM.\arabic{table}}
\renewcommand{\theequation}{SM.\arabic{equation}}
\setcounter{page}{1}
\renewcommand{\thepage}{SM \arabic{page}}

\section{Additional analyses of SKBD}\label{sec:addSKBD}

\subsection{Scenarios}\label{sec:scenarios}

\subsubsection{Fixed scenarios}

Table~\ref{tab:toxYan} lists the 20 fixed dose--toxicity scenarios considered in \citet{Yan2017Keyboard}. Each scenario involves five prespecified dose levels. The target DLT rate is $\phi=0.2$ for Scenarios~1--10 and $\phi=0.3$ for Scenarios~11--20. Across these scenarios, the true MTD ranges from the lowest to the highest dose level.

\begin{table}[!t]
	\centering
	\renewcommand{\arraystretch}{1.2}
	\caption{Twenty fixed dose--toxicity scenarios from \citet{Yan2017Keyboard} for five prespecified dose levels. The target DLT rate is $\phi=0.2$ for Scenarios~1--10 and $\phi=0.3$ for Scenarios~11--20. Across the 20 scenarios, the true MTD spans dose levels 1--5.}
	\begin{tabular}{c ccccc c c ccccc}
		\toprule
		& \multicolumn{5}{c}{  $\phi=0.2$} & & &
		\multicolumn{5}{c}{  $\phi=0.3$} \\
		\midrule
		Scenario & $d_1$ & $d_2$ & $d_3$ & $d_4$ & $d_5$ & &
		Scenario & $d_1$ & $d_2$ & $d_3$ & $d_4$ & $d_5$ \\
		1  & {\bf 0.20} & 0.26 & 0.40 & 0.45 & 0.46 & & 11 & {\bf 0.30} & 0.36 & 0.42 & 0.45 & 0.46 \\
		2  & {\bf 0.20} & 0.29 & 0.35 & 0.50 & 0.58 & & 12 & {\bf 0.30} & 0.40 & 0.55 & 0.60 & 0.70 \\
		3  & 0.10 & {\bf 0.20} & 0.25 & 0.35 & 0.40 & & 13 & 0.08 & {\bf 0.30} & 0.38 & 0.42 & 0.52 \\
		4  & 0.08 & {\bf 0.20} & 0.30 & 0.45 & 0.65 & & 14 & 0.13 & {\bf 0.30} & 0.42 & 0.50 & 0.80 \\
		5  & 0.04 & 0.06 & {\bf 0.20} & 0.32 & 0.50 & & 15 & 0.04 & 0.07 & {\bf 0.30} & 0.35 & 0.42 \\
		6  & 0.01 & 0.10 & {\bf 0.20} & 0.26 & 0.35 & & 16 & 0.01 & 0.12 & {\bf 0.30} & 0.41 & 0.55 \\
		7  & 0.05 & 0.06 & 0.07 & {\bf 0.20} & 0.31 & & 17 & 0.06 & 0.07 & 0.12 & {\bf 0.30} & 0.40 \\
		8  & 0.02 & 0.04 & 0.10 & {\bf 0.20} & 0.25 & & 18 & 0.02 & 0.05 & 0.16 & {\bf 0.30} & 0.36 \\
		9  & 0.01 & 0.02 & 0.07 & 0.08 & {\bf 0.20} & & 19 & 0.01 & 0.02 & 0.04 & 0.06 & {\bf 0.30} \\
		10 & 0.01 & 0.02 & 0.03 & 0.04 & {\bf 0.20} & & 20 & 0.06 & 0.07 & 0.08 & 0.12 & {\bf 0.30} \\
		\bottomrule
	\end{tabular}
	\label{tab:toxYan}
\end{table}

\subsubsection{Random scenarios}

Given a target DLT rate $\phi$ and $J$ prespecified dose levels, the algorithm proceeds as follows:

\begin{enumerate}[label=(\roman*), itemsep=2pt, topsep=0pt, parsep=0pt]
	\item Select one dose level $j \in \{1,\ldots,J\}$ with probability $1/J$ and treat it as the putative MTD location.
	
	\item Sample $ M \sim \mathrm{Beta}(\max\{J-j,0.5\},1)$, and define the upper bound of the toxicity probabilities as $B=\phi+(1-\phi)M$.
	
	\item Repeatedly generate $J$ monotone toxicity probabilities on $[0,B]$ until a scenario is obtained in which dose level $j$ is the MTD.
\end{enumerate}

\begin{figure}[!t]
	\centering
	\includegraphics[width=0.9\linewidth]{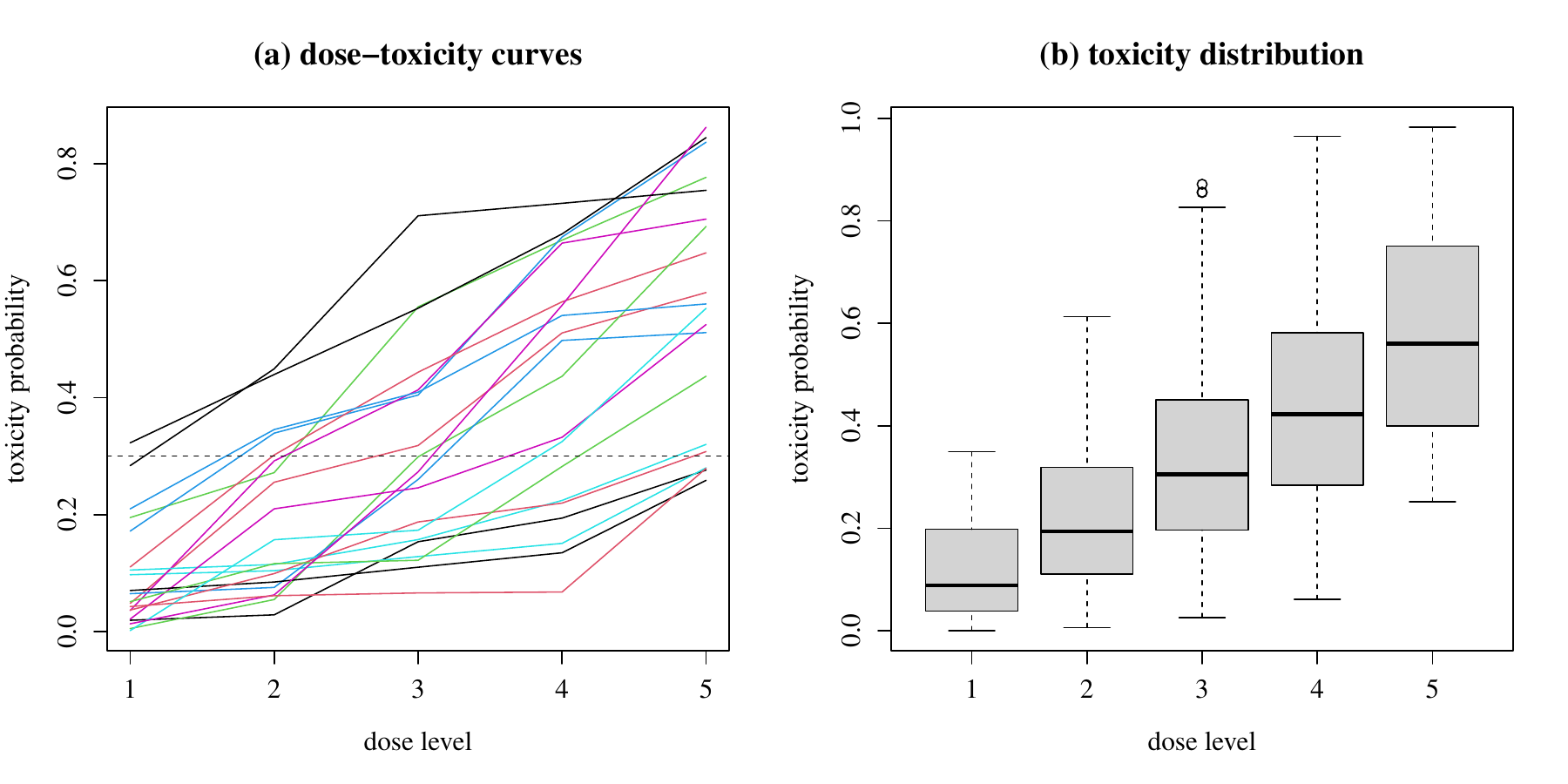}
	\caption{Randomly generated dose--toxicity scenarios for $J=5$ and $\phi=0.3$. (a): 20 representative randomly selected admissible monotone dose--toxicity curves, (b): the distribution of toxicity probabilities at each dose level across all 1,000 generated scenarios.}
	\label{fig:random-scenarios}
\end{figure}

For each combination of $J$ and $\phi$ considered in the simulation study, we generated 1,000 admissible random scenarios. The true MTD was therefore equally likely to occur at any of the $J$ prespecified dose levels. To ensure that the generated scenarios were clinically meaningful and suitable for evaluating MTD selection, we retained only scenarios satisfying the following conditions: (i) the true DLT probability at the MTD lay in $[\phi-0.05,\phi+0.05]$; (ii) the MTD was uniquely defined as the dose whose toxicity probability was closest to $\phi$; and (iii) the toxicity increments between the MTD and its adjacent doses, when such doses existed, were required to be at least $\varepsilon_1$ and $\varepsilon_2$ on the lower- and upper-dose sides, respectively, and no greater than 0.3.

These restrictions were imposed to avoid degenerate scenarios in which the target dose was poorly separated from neighbouring doses or in which the dose--toxicity curve changed implausibly abruptly. The generation procedure was repeated until 1,000 admissible scenarios had been collected for each combination of $J$ and $\phi$. Figure~\ref{fig:random-scenarios} displays a representative subset of the generated scenarios and the corresponding distribution of toxicity probabilities across dose levels.

\begin{table}[!t]
	\centering
	\vspace{0.5cm}
	\begin{minipage}{\textwidth}
		\centering
		\renewcommand{\arraystretch}{1.2}
		\caption{Accuracy performance under fixed scenarios. A higher value is better. All values are reported in percentage (\%).}
		\label{tab:fixed-accuracy}
		\resizebox{\textwidth}{!}{%
			\begin{tabular}{lccccccccccc}
				\toprule
				\itshape PCS & sc 1 & sc 2 & sc 3 & sc 4 & sc 5 & sc 6 & sc 7 & sc 8 & sc 9 & sc 10 & ave \\
				Keyboard & 49.2 & 53.2 & 38.5 & 49.3 & 53.6 & 39.3 & 45.5 & 37.1 & 59.1 & 72.8 & 49.8 \\
				SKBD     & 47.8 & 50.7 & 43.6 & 55.2 & 59.3 & 44.7 & 52.8 & 44.9 & 66.1 & 80.1 & 54.5 \\
				\hdashline
				& sc 11 & sc 12 & sc 13 & sc 14 & sc 15 & sc 16 & sc 17 & sc 18 & sc 19 & sc 20 & ave \\
				Keyboard & 46.6 & 55.3 & 50.5 & 54.5 & 46.8 & 53.1 & 51.7 & 43.7 & 83.5 & 73.9 & 56.0 \\
				SKBD     & 49.1 & 56.7 & 57.3 & 60.0 & 53.8 & 58.1 & 55.0 & 45.4 & 90.6 & 79.5 & 60.5 \\
				\midrule
				\itshape PCA & sc 1 & sc 2 & sc 3 & sc 4 & sc 5 & sc 6 & sc 7 & sc 8 & sc 9 & sc 10 & ave \\
				Keyboard & 64.2 & 67.3 & 34.4 & 39.3 & 35.0 & 29.1 & 26.2 & 23.9 & 30.1 & 37.3 & 38.7 \\
				SKBD     & 68.9 & 70.8 & 33.1 & 38.3 & 34.7 & 29.1 & 25.5 & 23.4 & 31.8 & 39.0 & 39.5 \\
				\hdashline
				& sc 11 & sc 12 & sc 13 & sc 14 & sc 15 & sc 16 & sc 17 & sc 18 & sc 19 & sc 20 & ave \\
				Keyboard & 59.9 & 64.6 & 42.6 & 43.7 & 36.8 & 38.1 & 31.3 & 29.1 & 45.9 & 37.8 & 43.0 \\
				SKBD     & 66.5 & 72.0 & 42.7 & 41.3 & 36.9 & 37.2 & 29.8 & 27.5 & 45.6 & 37.4 & 43.7 \\
				\bottomrule
			\end{tabular}
		}
		\begin{flushleft}
			\footnotesize PCS = percentage of correct selection; PCA = percentage of correct allocation.
		\end{flushleft}
	\end{minipage}
	
	\vspace{1cm}
	
	\begin{minipage}{\textwidth}
		\centering
		\renewcommand{\arraystretch}{1.2}
		\caption{Safety performance under fixed scenarios. A lower value is better. All values are reported in percentage (\%).}
		\label{tab:fixed-safety}
		\resizebox{\textwidth}{!}{%
			\begin{tabular}{lccccccccccc}
				\toprule
				\itshape Above-MTD & sc 1 & sc 2 & sc 3 & sc 4 & sc 5 & sc 6 & sc 7 & sc 8 & sc 9 & sc 10 & ave \\
				Keyboard & 35.8 & 32.7 & 6.8 & 21.3 & 17.1 & 19.0 & 12.4 & 0.0 & 0.0 & 0.0 & 14.5 \\
				SKBD     & 31.1 & 29.2 & 7.2 & 19.9 & 16.9 & 19.5 & 13.4 & 0.0 & 0.0 & 0.0 & 13.7 \\
				\hdashline
				& sc 11 & sc 12 & sc 13 & sc 14 & sc 15 & sc 16 & sc 17 & sc 18 & sc 19 & sc 20 & ave \\
				Keyboard & 40.1 & 35.4 & 29.0 & 24.2 & 7.7 & 21.9 & 17.5 & 17.9 & 0.0 & 0.0 & 19.4 \\
				SKBD     & 33.5 & 28.0 & 26.0 & 21.2 & 7.5 & 19.7 & 16.8 & 17.4 & 0.0 & 0.0 & 17.0 \\
				\midrule
				\itshape ROD & sc 1 & sc 2 & sc 3 & sc 4 & sc 5 & sc 6 & sc 7 & sc 8 & sc 9 & sc 10 & ave \\
				Keyboard & 22.1 & 18.3 & 1.6 & 9.1 & 4.5 & 8.7 & 0.0 & 0.0 & 0.0 & 0.0 & 6.4 \\
				SKBD     & 14.3 & 12.0 & 1.0 & 5.2 & 2.6 & 6.5 & 0.0 & 0.0 & 0.0 & 0.0 & 4.2 \\
				\hdashline
				& sc 11 & sc 12 & sc 13 & sc 14 & sc 15 & sc 16 & sc 17 & sc 18 & sc 19 & sc 20 & ave \\
				Keyboard & 29.5 & 22.4 & 18.1 & 11.3 & 0.0 & 6.0 & 0.0 & 0.0 & 0.0 & 0.0 & 8.7 \\
				SKBD     & 20.4 & 11.8 & 12.4 & 6.4 & 0.0 & 3.1 & 0.0 & 0.0 & 0.0 & 0.0 & 5.4 \\
				\bottomrule
			\end{tabular}
		}
		\begin{flushleft}
			\footnotesize Above-MTD = proportion of patients treated above the MTD; ROD = risk of overdosing.
		\end{flushleft}
	\end{minipage}
	\vspace{1cm}
\end{table}

\subsection{Scenario-specific results for the fixed scenarios}

Tables~\ref{tab:fixed-accuracy} and~\ref{tab:fixed-safety} provide the scenario-specific values corresponding to Figures~5 and~6 in the main text. Table~\ref{tab:fixed-accuracy} reports the two accuracy measures, PCS and PCA, whereas Table~\ref{tab:fixed-safety} reports the two safety measures, above-MTD and ROD. 

\subsection{Additional analyses on kernel specification}

\subsubsection{Asymmetric versus symmetric kernels}\label{sec:asyVSsy}

\begin{table}[!t]
	\centering
	\renewcommand{\arraystretch}{1.2}
	\caption{Comparison of asymmetric and symmetric kernels under fixed scenarios. All values are reported in percentage (\%).}
	\label{tab:asym-vs-sym}
	\begin{tabular}{lcccc}
		\toprule
		& \multicolumn{2}{c}{Accuracy} & \multicolumn{2}{c}{Safety} \\
		\cmidrule(lr){2-3} \cmidrule(lr){4-5}
		& PCS & PCA & Above-MTD & ROD \\
		\midrule
		\multicolumn{5}{c}{Scenario 1--10} \\
		Asymmetric       & 54.5 & 39.5 & 13.7 & 4.2 \\
		Symmetric (0.8) & 51.8 & 39.6 & 21.3 & 9.1 \\
		Symmetric (0.5) & 52.9 & 40.3 & 20.2 & 9.1 \\
		Symmetric (0.2) & 53.6 & 39.8 & 16.9 & 6.9 \\
		\midrule
		\multicolumn{5}{c}{Scenario 11--20} \\
		Asymmetric       & 60.5 & 43.7 & 17.0 & 5.4 \\
		Symmetric (0.8) & 55.7 & 41.1 & 27.6 & 12.1 \\
		Symmetric (0.5) & 57.6 & 42.3 & 25.2 & 11.2 \\
		Symmetric (0.2) & 58.6 & 43.3 & 19.9 & 8.2 \\
		\bottomrule
	\end{tabular}
\end{table}

Table~\ref{tab:asym-vs-sym} compares the proposed asymmetric kernel with several symmetric kernels within the SKBD framework. For the symmetric kernels, the values in parentheses denote the common nearest-neighbour kernel value $k_{\mathrm{sym}}(d_j,d_j\pm\sigma)$ after dose standardisation. Thus, all specifications are compared under the same nearest-neighbour calibration scheme as that used in the main text.

Across both sets of fixed scenarios, the asymmetric kernel provided the most favourable overall operating characteristics. For Scenarios~1--10, it achieved the highest PCS (54.5\%), compared with 51.8\%, 52.9\%, and 53.6\% under the three symmetric kernels, while PCA was broadly similar across specifications (39.5\% versus 39.6\%--40.3\%). The safety advantage was more pronounced: above-MTD decreased to 13.7\% under the asymmetric kernel, compared with 21.3\%, 20.2\%, and 16.9\% under the symmetric kernels, and ROD decreased to 4.2\%, compared with 9.1\%, 9.1\%, and 6.9\%, respectively.

A similar pattern was observed for Scenarios~11--20. The asymmetric kernel again yielded the highest PCS (60.5\%) and PCA (43.7\%), while also achieving the lowest above-MTD (17.0\%) and ROD (5.4\%). In comparison, the symmetric kernels produced PCS values between 55.7\% and 58.6\%, PCA values between 41.1\% and 43.3\%, above-MTD values between 19.9\% and 27.6\%, and ROD values between 8.2\% and 12.1\%.

Overall, these results indicate that the asymmetric kernel improves safety substantially relative to the symmetric alternatives, while maintaining---and in most settings improving---accuracy. This supports the use of asymmetric borrowing as the default specification in SKBD.

\subsubsection{Sensitivity to nearest-neighbour kernel specification}\label{sec:senKernel}

We conducted a sensitivity analysis to assess the impact of the nearest-neighbour kernel specification on SKBD performance. Assuming equally spaced standardised dose levels, we varied the kernel values assigned to the nearest lower and higher neighbours. Specifically, the value for the lower neighbour, $k(d_j,d_{j-1})$, ranged from 0.1 to 0.3, whereas that for the higher neighbour, $k(d_j,d_{j+1})$, ranged from 0.7 to 0.9.

\begin{table}[!t]
	\centering
	\renewcommand{\arraystretch}{1.2}
	\caption{Sensitivity analysis of SKBD under different nearest-neighbour kernel specifications. All values are reported in percentage (\%). Boldface indicates the default specification, $(k(d_j,d_{j-1}),k(d_j,d_{j+1}))=(0.2,0.8)$, used in the main text.}
	\label{tab:sensitive}
	\begin{tabular}{cccccc}
		\toprule
		& & \multicolumn{2}{c}{Accuracy} & \multicolumn{2}{c}{Safety} \\
		\cmidrule(lr){3-4} \cmidrule(lr){5-6}
		$k(d_j,d_{j-1})$ & $k(d_j,d_{j+1})$ & PCS & PCA & Above-MTD & ROD \\
		\midrule
		\multicolumn{6}{c}{Scenario 1--10} \\
		& 0.7 & 52.3 & 39.1 & 12.8 & 4.3 \\
		0.1 & 0.8 & 52.4 & 38.8 & 12.6 & 4.1 \\
		& 0.9 & 52.4 & 38.5 & 12.2 & 3.6 \\
		\hdashline
		& 0.7 & 54.5 & 39.7 & 14.2 & 4.7 \\
		\textbf{0.2} & \textbf{0.8} & \textbf{54.5} & \textbf{39.5} & \textbf{13.7} & \textbf{4.2} \\
		& 0.9 & 54.5 & 39.1 & 13.3 & 3.7 \\
		\hdashline
		& 0.7 & 54.2 & 39.8 & 14.9 & 5.1 \\
		0.3 & 0.8 & 54.3 & 39.6 & 14.5 & 4.6 \\
		& 0.9 & 54.4 & 39.1 & 13.9 & 4.2 \\
		\midrule
		\multicolumn{6}{c}{Scenario 11--20} \\
		& 0.7 & 59.9 & 43.7 & 17.2 & 5.6 \\
		0.1 & 0.8 & 60.3 & 43.5 & 16.8 & 5.3 \\
		& 0.9 & 60.8 & 43.2 & 16.4 & 5.0 \\
		\hdashline
		& 0.7 & 60.2 & 43.8 & 17.5 & 5.9 \\
		\textbf{0.2} & \textbf{0.8} & \textbf{60.5} & \textbf{43.7} & \textbf{17.0} & \textbf{5.4} \\
		& 0.9 & 61.2 & 43.3 & 16.6 & 5.0 \\
		\hdashline
		& 0.7 & 60.0 & 44.2 & 17.9 & 6.3 \\
		0.3 & 0.8 & 60.4 & 44.1 & 17.4 & 5.7 \\
		& 0.9 & 60.8 & 43.7 & 17.0 & 5.2 \\
		\bottomrule
	\end{tabular}
\end{table}

Table~\ref{tab:sensitive} summarises the results across Scenarios~1--20. Overall, SKBD performance was fairly robust to these changes, with only moderate variation in both accuracy and safety across the range of specifications considered. Increasing the kernel value for the higher neighbour tended to produce slightly higher PCS and lower overdose risk, at the cost of a small reduction in PCA. The default specification, $(k(d_j,d_{j-1}),k(d_j,d_{j+1}))=(0.2,0.8)$, did not uniformly optimise every operating characteristic, but it provided a favourable overall balance between accuracy and safety. 

\subsection{Sensitivity to cohort size}\label{sec:cohort1}

We conducted an additional sensitivity analysis to assess whether the operating characteristics of SKBD were affected by the choice of cohort size. 
Specifically, we repeated the fixed-scenario simulation study for both SKBD and the keyboard design using single-patient cohorts in place of the default cohort size of three. 
The maximum sample size remained 30 patients, and all other settings were kept unchanged from the main analysis. 

\begin{figure}[!t]
	\centering
	\includegraphics[width=0.8\linewidth]{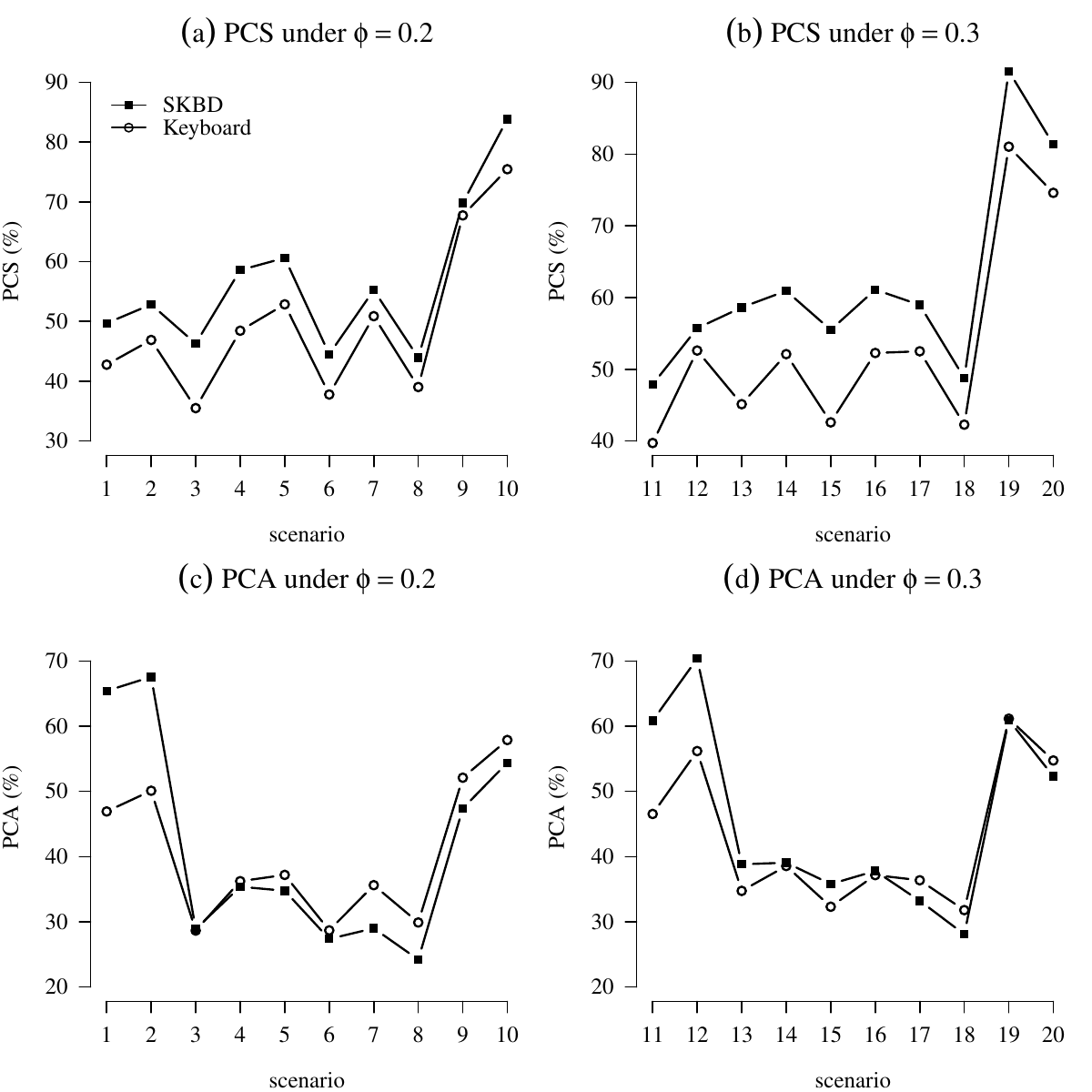}
	\caption{Accuracy performance of SKBD and keyboard design under fixed scenarios with cohort size 1. Panels (a) and (b) show PCS for $\phi=0.2$ and $\phi=0.3$, respectively; panels (c) and (d) show PCA. A higher value is better.}
	\label{fig:fixed-accuracy-cohortsize1}
\end{figure}

\begin{figure}[!t]
	\centering
	\includegraphics[width=0.8\linewidth]{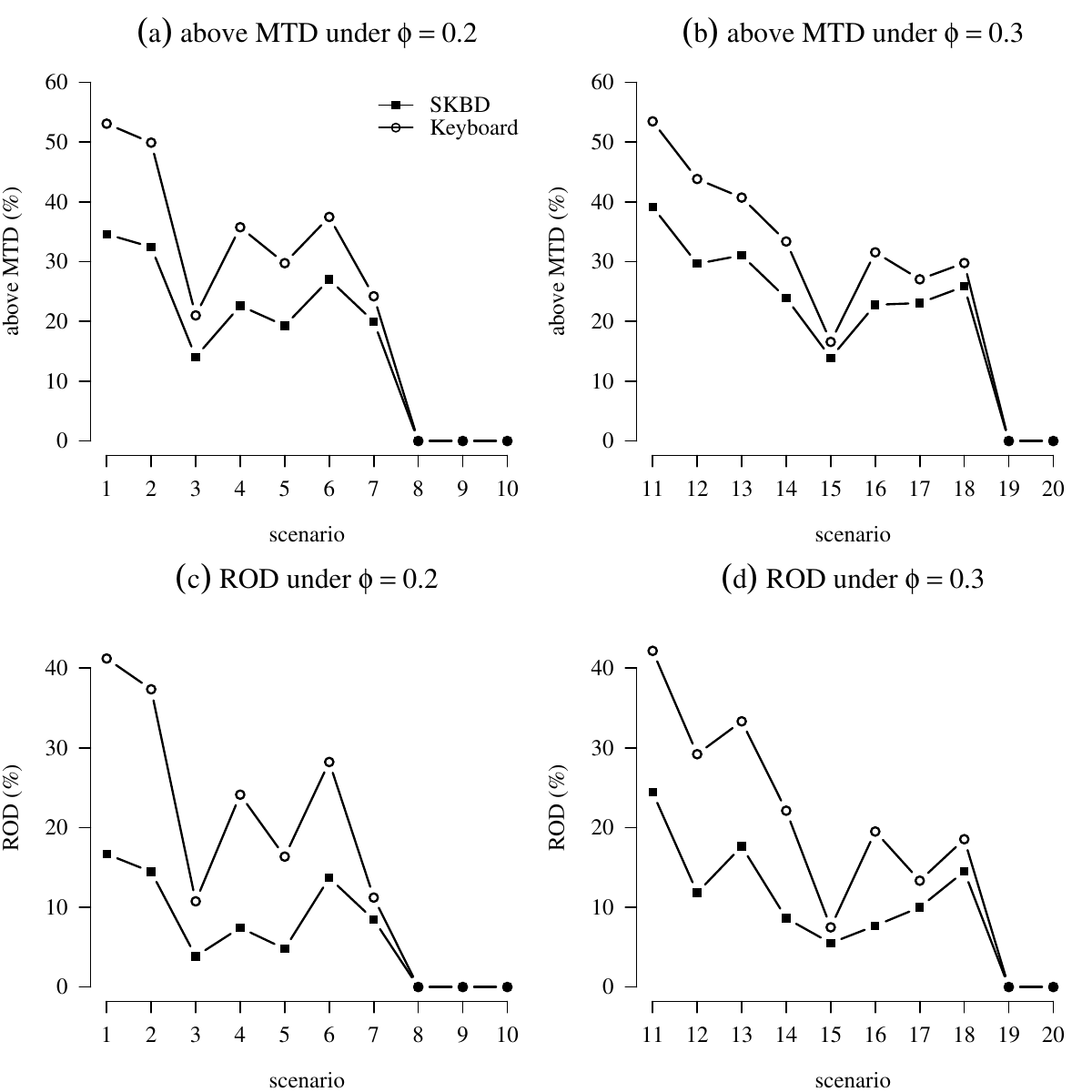}
	\caption{Safety performance of SKBD and keyboard design under fixed scenarios with cohort size 1. Panels (a) and (b) show above-MTD for $\phi=0.2$ and $\phi=0.3$, respectively; panels (c) and (d) show ROD. A lower value is better.}
	\label{fig:fixed-safety-cohortsize1}
\end{figure}

The resulting operating characteristics are shown in Figures~\ref{fig:fixed-accuracy-cohortsize1} and~\ref{fig:fixed-safety-cohortsize1}. 
Overall, the conclusions were broadly consistent with those obtained under the default cohort size. 
Relative to the keyboard design, SKBD continued to improve MTD-selection performance while maintaining a favourable safety profile under single-patient updating. 
These findings show that SKBD works well under varying cohort sizes.

\section{Details of TITE-SKBD}\label{sec:SM-TITE-SKBD}

\subsection{Methodological details}
We adopt the effective sample size (ESS) framework of \citet{Lin2020ESS} to incorporate pending DLT outcomes through \emph{effective} binomial statistics, and then embed these effective statistics into SKBD.

Suppose toxicities are assessed over a fixed period $(0,\tau)$, referred to as the DLT assessment window. 
The value of $\tau$ is determined in consultation with clinicians and chosen to be sufficiently long to capture clinically relevant toxicities (e.g., one chemotherapy cycle such as 21 or 28 days, or longer for agents with delayed effects).

Consider an interim decision time. 
For patient $i$, define the (latent) binary DLT outcome
\[
x_i =
\begin{cases}
	1, & \text{if a DLT occurs during } (0,\tau],\\
	0, & \text{otherwise},
\end{cases}
\]
and let $t_i$ be the actual time to DLT when $x_i=1$; otherwise set $t_i=\infty$.
Let $u_i\in[0,\tau]$ denote the observed follow-up time at the interim analysis.
If $u_i=\tau$, the patient has completed the assessment window; if $u_i<\tau$, the DLT outcome may still be pending. 
Define the ascertainment indicator
\[
\delta_i =
\begin{cases}
	1, & \text{if } x_i \text{ is ascertained at the interim time},\\
	0, & \text{otherwise}.
\end{cases}
\]
In particular, $\delta_i=1$ if a DLT has already occurred ($t_i\le u_i<\tau$), or if the patient has completed follow-up without DLT ($u_i=\tau<t_i=\infty$); otherwise, $\delta_i=0$.

To quantify the amount of information contributed by a patient with incomplete follow-up, we assign a weight $w_i\in[0,1]$.
If the outcome is fully ascertained ($\delta_i=1$), we set $w_i=1$.
If the outcome is pending ($\delta_i=0$), we define
\[
w_i=
\begin{cases}
	1, & \delta_i=1,\\
	\Pr(t_i\le u_i \mid x_i=1), & \delta_i=0,
\end{cases}
\]
so that longer DLT-free follow-up corresponds to a larger contribution at the interim time. 
Under the working assumption that, conditional on $x_i=1$, the DLT time is uniformly distributed over $(0,\tau]$, the weight simplifies to $w_i=u_i/\tau$.
Although the uniform assumption is simple, it has been shown to yield robust operating characteristics in a variety of dose-finding settings \citep{Cheung2000TITECRM, Yuan2018TITEBOIN, Lin2020ESS}.

Let $n_j$ be the number of patients treated at dose level $d_j$ up to the interim time.
Define the number of \emph{observed} DLTs at dose $d_j$ as
\[
\tilde y_j=\sum_{i=1}^{n_j} \delta_i x_i,
\]
and the number of fully observed non-DLT outcomes as
\[
m_j=\sum_{i=1}^{n_j} \delta_i(1-x_i).
\]
For patients with pending outcomes ($\delta_i=0$), we treat their accumulated DLT-free follow-up as contributing a fractional amount to the non-DLT count. 
Accordingly, the \emph{effective} number of non-DLTs is
\[
\tilde m_j = m_j+\sum_{i=1}^{n_j}(1-\delta_i)w_i,
\]
and the total \emph{effective} sample size is
\[
\tilde n_j=\tilde y_j+\tilde m_j.
\]
When all outcomes are fully ascertained, $\delta_i=1$ for all patients and thus $\tilde y_j=y_j$, $\tilde n_j=n_j$, recovering the complete-data binomial statistics.

The time-to-event shared keyboard design (TITE-SKBD) follows the same decision structure as SKBD, with one modification: 
at each interim decision time, we first construct the dose-specific effective data 
$\tilde{\mathcal{D}}=\{(\tilde n_j,\tilde y_j): j=1,\ldots,J\}$.
We then apply the same sharing step across doses as in SKBD to these effective dose-level statistics, and use the resulting shared effective data to update the posterior distribution of the toxicity probability $\pi_j$. 

For patient safety, we adopt the escalation suspension rule proposed by \citet{Lin2020ESS}, which prohibits dose escalation until at least two patients have completed the DLT assessment at the current dose level; alternative suspension rules have also been discussed in the literature (e.g., \citealp{Zhou2024TITE}).
Once this requirement is satisfied, the strongest-key identification and dose transition rules are applied exactly as in SKBD based on the current posterior distribution. 
This modification enables real-time dose assignment without delaying accrual, while retaining the ethical integrity and simplicity of the original shared keyboard decision rule.

\subsection{Operating characteristics}\label{sec:OC-TITE}

\begin{figure}[!t]
	\centering
	\includegraphics[width=0.8\linewidth]{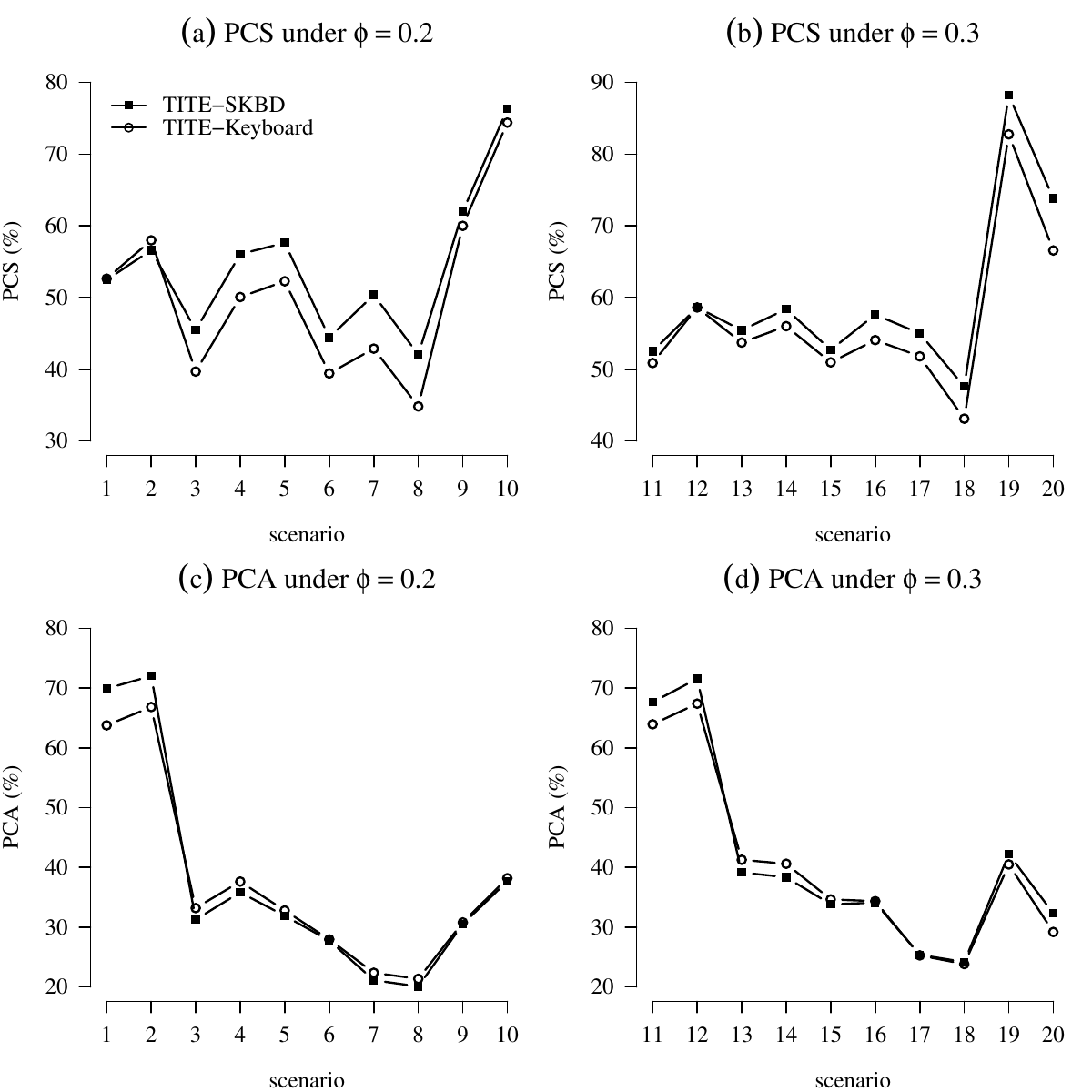}
	\caption{Accuracy performance of TITE-SKBD and TITE-Keyboard under fixed scenarios with late-onset toxicity. Panels (a) and (b) show PCS for $\phi=0.2$ and $\phi=0.3$, respectively; panels (c) and (d) show PCA. A higher value is better.}
	\label{fig:fixed-accuracy-tite}
\end{figure}

\begin{figure}[!t]
	\centering
	\includegraphics[width=0.8\linewidth]{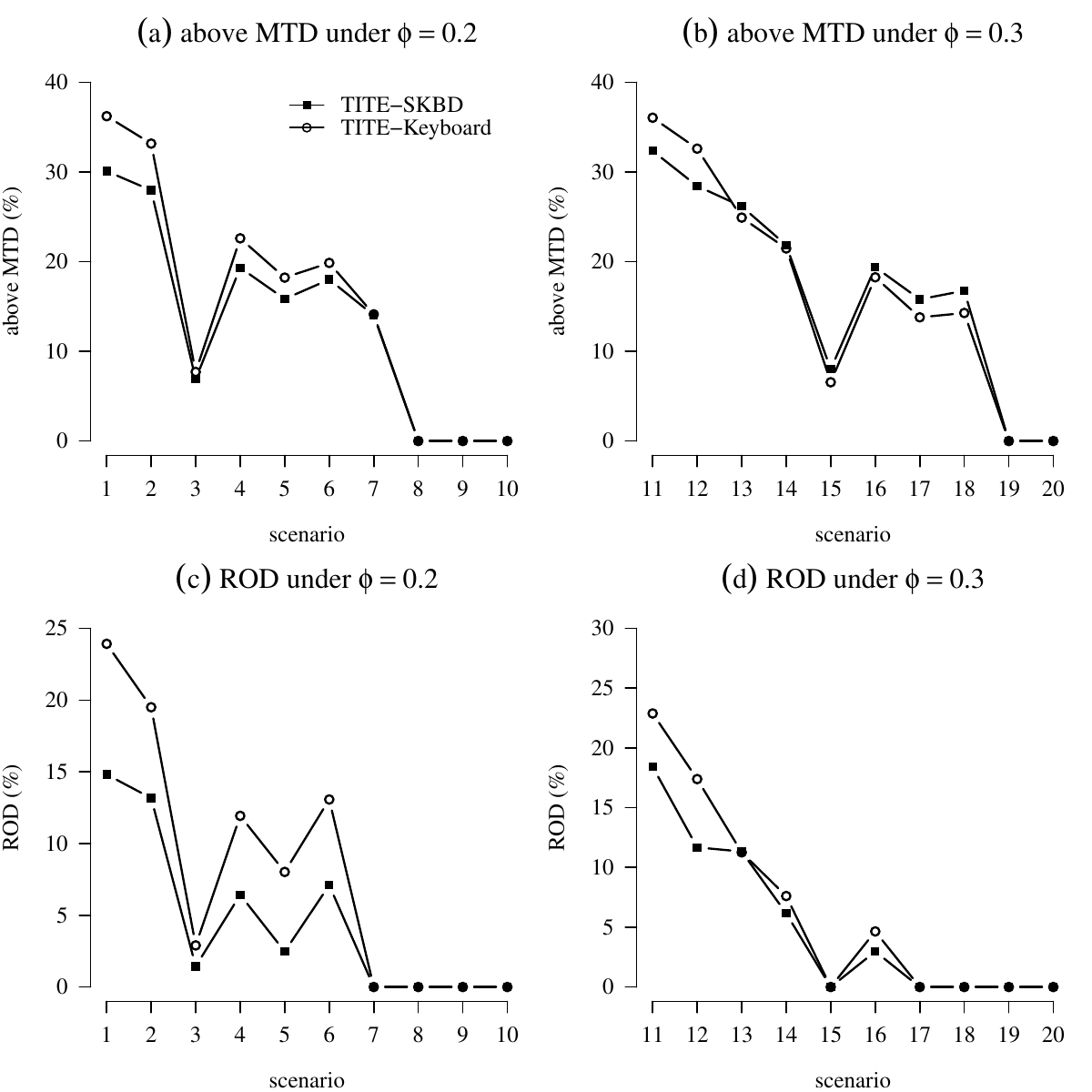}
	\caption{Safety performance of TITE-SKBD and TITE-Keyboard under fixed scenarios with late-onset toxicity. Panels (a) and (b) compare above-MTD for $\phi=0.2$ and $\phi=0.3$, respectively; panels (c) and (d) compare ROD. A lower value is better.}
	\label{fig:fixed-safety-tite}
\end{figure}

\begin{table}[!t]
	\centering
	\vspace{0.5cm}
	\begin{minipage}{\textwidth}
		\centering
		\renewcommand{\arraystretch}{1.2}
		\caption{Accuracy performance of TITE-Keyboard and TITE-SKBD under fixed scenarios. A higher value is better. All values are reported in percentage (\%).}
		\label{tab:fixed-accuracy-tite}
		\resizebox{\textwidth}{!}{%
			\begin{tabular}{lccccccccccc}
				\toprule
				\itshape PCS & sc 1 & sc 2 & sc 3 & sc 4 & sc 5 & sc 6 & sc 7 & sc 8 & sc 9 & sc 10 & ave \\
				TITE-Keyboard & 52.6 & 58.0 & 39.7 & 50.1 & 52.3 & 39.4 & 42.9 & 34.8 & 60.0 & 74.4 & 50.4 \\
				TITE-SKBD     & 52.5 & 56.6 & 45.5 & 56.0 & 57.6 & 44.4 & 50.4 & 42.0 & 62.0 & 76.3 & 54.3 \\
				\hdashline
				& sc 11 & sc 12 & sc 13 & sc 14 & sc 15 & sc 16 & sc 17 & sc 18 & sc 19 & sc 20 & ave \\
				TITE-Keyboard & 50.9 & 58.6 & 53.7 & 56.0 & 51.0 & 54.1 & 51.8 & 43.1 & 82.8 & 66.6 & 56.8 \\
				TITE-SKBD     & 52.5 & 58.6 & 55.4 & 58.4 & 52.7 & 57.6 & 55.0 & 47.6 & 88.2 & 73.8 & 60.0 \\
				\midrule
				\itshape PCA & sc 1 & sc 2 & sc 3 & sc 4 & sc 5 & sc 6 & sc 7 & sc 8 & sc 9 & sc 10 & ave \\
				TITE-Keyboard & 63.8 & 66.8 & 33.2 & 37.6 & 32.8 & 28.0 & 22.4 & 21.4 & 30.8 & 38.2 & 37.5 \\
				TITE-SKBD     & 69.9 & 72.0 & 31.3 & 35.8 & 31.9 & 27.8 & 21.1 & 20.1 & 30.5 & 37.7 & 37.8 \\
				\hdashline
				& sc 11 & sc 12 & sc 13 & sc 14 & sc 15 & sc 16 & sc 17 & sc 18 & sc 19 & sc 20 & ave \\
				TITE-Keyboard &63.9 & 67.4 & 41.3 & 40.6 & 34.7 & 34.4 & 25.3 & 23.8 & 40.5 & 29.2 & 40.1 \\
				TITE-SKBD     & 67.6 & 71.6 & 39.2 & 38.4 & 33.8 & 34.1 & 25.4 & 24.2 & 42.2 & 32.3 & 40.9 \\
				\bottomrule
			\end{tabular}
		}
	\end{minipage}
	
	\vspace{1cm}
	
	\begin{minipage}{\textwidth}
		\centering
		\renewcommand{\arraystretch}{1.2}
		\caption{Safety performance under fixed scenarios. A lower value is better. All values are reported in percentage (\%).}
		\label{tab:fixed-safety-tite}
		\resizebox{\textwidth}{!}{%
			\begin{tabular}{lccccccccccc}
				\toprule
				\itshape Above-MTD & sc 1 & sc 2 & sc 3 & sc 4 & sc 5 & sc 6 & sc 7 & sc 8 & sc 9 & sc 10 & ave \\
				TITE-Keyboard & 36.2 & 33.2 & 7.7 & 22.6 & 18.2 & 19.9 & 14.1 & 0.0 & 0.0 & 0.0 & 15.2 \\
				TITE-SKBD     & 30.1 & 28.0 & 6.9 & 19.3 & 15.8 & 18.0 & 14.0 & 0.0 & 0.0 & 0.0 & 13.2 \\
				\hdashline
				& sc 11 & sc 12 & sc 13 & sc 14 & sc 15 & sc 16 & sc 17 & sc 18 & sc 19 & sc 20 & ave \\
				TITE-Keyboard & 36.1 & 32.6 & 24.9 & 21.5 & 6.6 & 18.3 & 13.8 & 14.3 & 0.0 & 0.0 & 16.8 \\
				TITE-SKBD     & 32.4 & 28.4 & 26.2 & 21.8 & 8.0 & 19.4 & 15.8 & 16.8 & 0.0 & 0.0 & 16.9 \\
				\midrule
				\itshape ROD & sc 1 & sc 2 & sc 3 & sc 4 & sc 5 & sc 6 & sc 7 & sc 8 & sc 9 & sc 10 & ave \\
				TITE-Keyboard & 23.9 & 19.5 & 2.9 & 11.9 & 8.0 & 13.1 & 0.0 & 0.0 & 0.0 & 0.0 & 7.9 \\
				TITE-SKBD     & 14.8 & 13.2 & 1.4 & 6.4 & 2.5 & 7.1 & 0.0 & 0.0 & 0.0 & 0.0 & 4.5 \\
				\hdashline
				& sc 11 & sc 12 & sc 13 & sc 14 & sc 15 & sc 16 & sc 17 & sc 18 & sc 19 & sc 20 & ave \\
				TITE-Keyboard & 22.9 & 17.4 & 11.3 & 7.6 & 0.0 & 4.6 & 0.0 & 0.0 & 0.0 & 0.0 & 6.4 \\
				TITE-SKBD     & 18.4 & 11.7 & 11.3 & 6.2 & 0.0 & 3.0 & 0.0 & 0.0 & 0.0 & 0.0 & 5.1 \\
				\bottomrule
			\end{tabular}
		}
	\end{minipage}
	\vspace{1cm}
\end{table}

We conducted simulation studies to evaluate the operating characteristics of TITE-SKBD under late-onset toxicity. 
Following the fixed-scenario study for the complete-data setting, we considered the same 20 scenarios in Table~\ref{tab:toxYan}. 
Patients were enrolled in cohorts of three, with a maximum sample size of 30. 
The DLT assessment window was set to $\tau=3$ months, and patients were enrolled at a rate of two patients per month. 
We compared the proposed TITE-SKBD with the standard TITE-Keyboard design. (Both were implemented using the \texttt{SKBD} package.) The results are summarised in Figures~\ref{fig:fixed-accuracy-tite}--\ref{fig:fixed-safety-tite} and Tables~\ref{tab:fixed-accuracy-tite}--\ref{tab:fixed-safety-tite}.

Overall, the late-onset extension showed a pattern broadly consistent with that observed for the ordinary SKBD design. In particular, TITE-SKBD improved MTD-selection performance over TITE-Keyboard across both target toxicity settings. As shown in Figure~\ref{fig:fixed-accuracy-tite} and Table~\ref{tab:fixed-accuracy-tite}, the average PCS increased from 50.4\% to 54.3\% when $\phi=0.2$ and from 56.8\% to 60.0\% when $\phi=0.3$. The average PCA also increased, although more modestly, from 37.5\% to 37.8\% for $\phi=0.2$ and from 40.1\% to 40.9\% for $\phi=0.3$. Thus, as in the complete-data setting, incorporating shared information across neighbouring doses improved final MTD identification while maintaining comparable or slightly improved patient allocation to the true MTD.

The safety results are reported in Figure~\ref{fig:fixed-safety-tite} and Table~\ref{tab:fixed-safety-tite}. For the low-target scenarios ($\phi=0.2$), TITE-SKBD yielded a more favourable safety profile than TITE-Keyboard, with the average above-MTD allocation decreasing from 15.2\% to 13.2\% and the average risk of overdosing (ROD) decreasing from 7.9\% to 4.5\%. For the high-target scenarios ($\phi=0.3$), the average above-MTD allocation was essentially unchanged (16.8\% under TITE-Keyboard versus 16.9\% under TITE-SKBD), whereas the average ROD decreased from 6.4\% to 5.1\%. 
These results indicate that the safety advantage of TITE-SKBD is more evident for ROD than for above-MTD allocation. For $\phi=0.2$, both safety measures decreased under TITE-SKBD, whereas for $\phi=0.3$, the above-MTD allocation remained comparable and ROD was reduced. 

These results show that TITE-SKBD preserves the main advantage of the ordinary SKBD design in the presence of pending DLT outcomes. Relative to the complete-data SKBD results, the safety advantage under late-onset toxicity is somewhat attenuated, particularly for above-MTD allocation when $\phi=0.3$, but the overall qualitative pattern remains the same. This supports the use of the shared, overdose-aware borrowing mechanism in trials where interim decisions must be made before all toxicity outcomes are fully observed.

\section{Software interface}\label{sec:software}

\begin{figure}[!t]
	\centering
	\includegraphics[width=\linewidth]{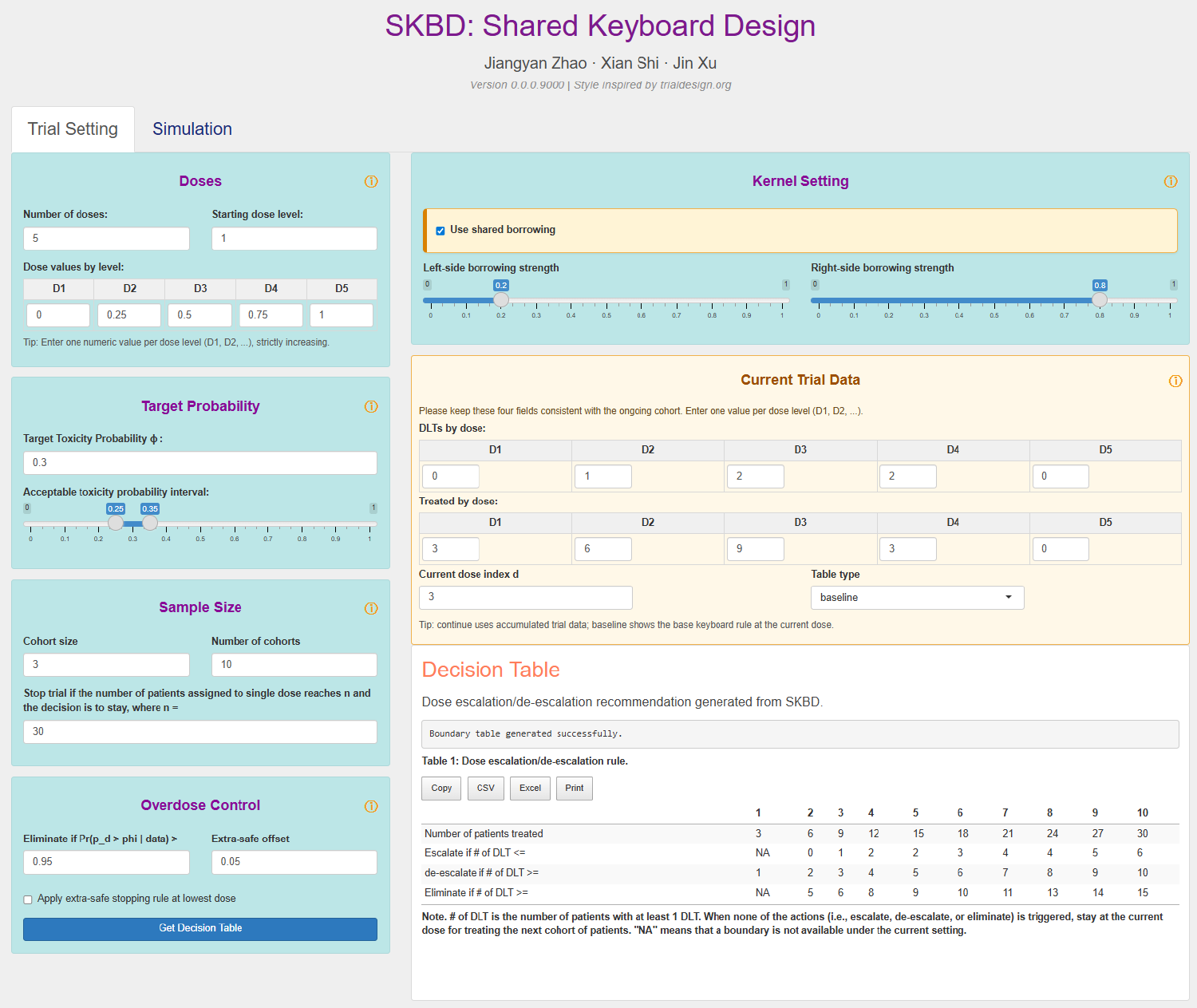}
	\caption{Trial-setting and decision-table interface of the Shiny-based graphical user interface for the \texttt{SKBD} package. Information icons in the input panels provide short explanations of the corresponding design components.}
	\label{fig:shiny-setting}
\end{figure}

\begin{figure}[!t]
	\centering
	\includegraphics[width=\linewidth]{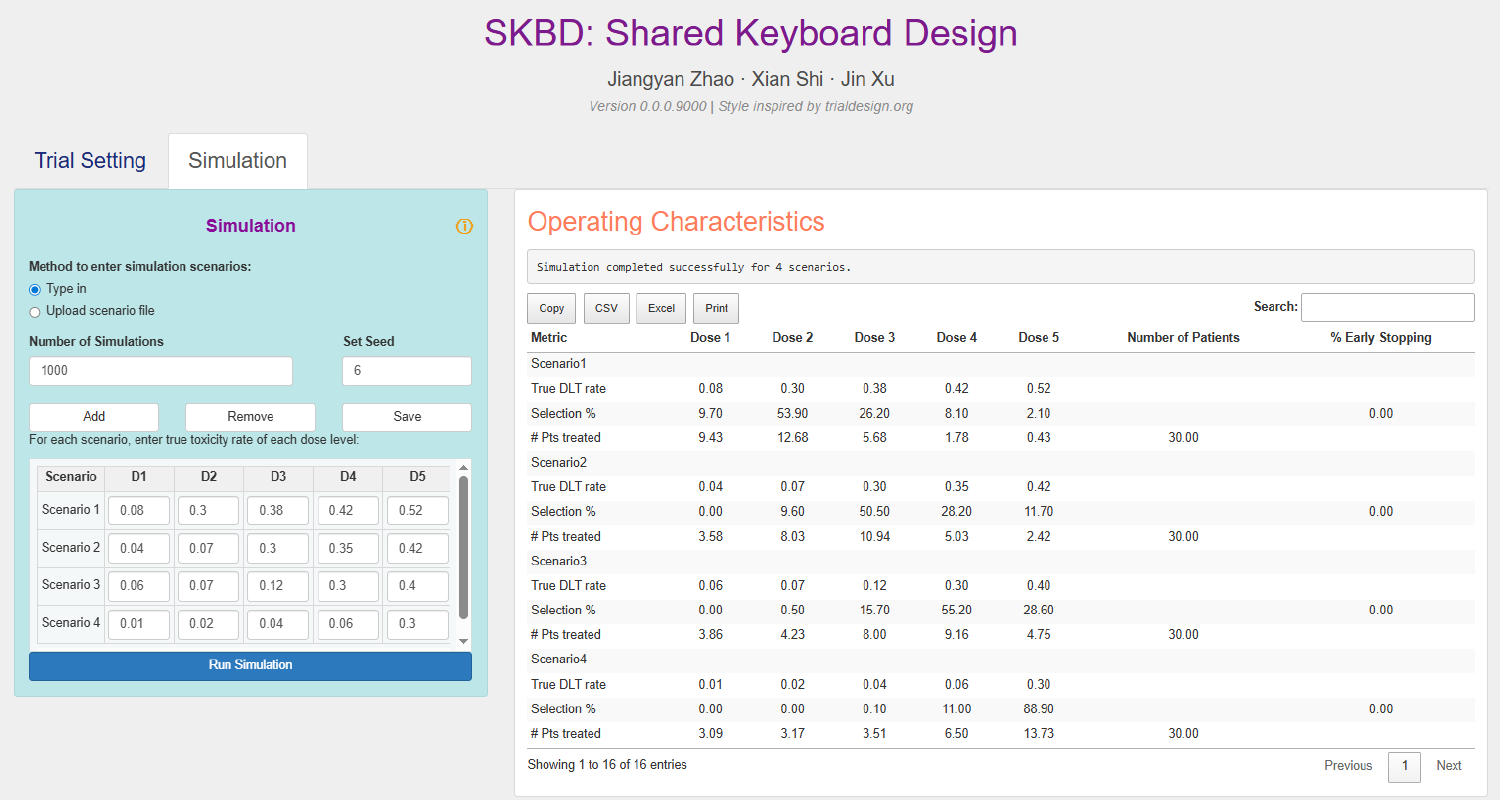}
	\caption{Simulation and operating-characteristics interface of the Shiny-based graphical user interface for the \texttt{SKBD} package.}
	\label{fig:shiny-simulation}
\end{figure}

To support implementation of the proposed designs, we developed a Shiny-based graphical user interface for the \texttt{SKBD} package. 
The interface provides two main modules and adopts a layout style of designs at \url{https://www.trialdesign.org}. 
The module allows users to specify dose levels, the target toxicity probability, sample size, overdose-control rules, shared-borrowing settings, and current trial data, and then generates the corresponding dose-assignment decision table. 
The simulation module allows users to input dose--toxicity scenarios, run simulation studies, and summarise the operating characteristics of the design. 
Short explanatory notes are provided through the information icons in the upper-right corner of the main input panels, so that users can check the meaning of each module and key parameter while using the interface. 
Figures~\ref{fig:shiny-setting} and \ref{fig:shiny-simulation} show representative screenshots of these two modules.

\end{document}